\documentclass[runningheads,envcountsame]{llncs}
\usepackage[T1]{fontenc}
\usepackage{microtype}
\usepackage{graphicx}  
\usepackage{appendix}
\usepackage{amsmath}
\usepackage{amssymb}
\usepackage{tikz}
\usetikzlibrary{calc}
\usepackage{hyperref}
\usepackage{subcaption}

\spnewtheorem{observation}[theorem]{Observation}{\bfseries}{\itshape}

\newcommand{\conv}{\operatorname{conv}}
\renewcommand{\succ}{\operatorname{succ}}
\newcommand{\pred}{\operatorname{pred}}
\newcommand{\ccw}{\operatorname{ccw}}
\newcommand{\cw}{\operatorname{cw}}
\newcommand{\inc}{\operatorname{in}}
\newcommand{\out}{\operatorname{out}}
\title{Quadratic Reconfiguration of Plane Spanning Paths on Two-Layer Point Sets}
\titlerunning{Quadratic Reconfiguration of Plane Spanning Paths}
\author{Samuel Avril\inst{1}\orcidID{0009-0007-2333-6825} \and
Yan Gerard\inst{2}\orcidID{0000-0002-2664-0650} \and
Guilherme D. da Fonseca\inst{3}\orcidID{0000-0002-9807-028X}}
\authorrunning{S. Avril et al.}
\institute{ENS de Lyon, Lyon, France\\
\email{samuel.avril@ens-lyon.fr}\and
LIMOS, Universit\'e Clermont Auvergne, Clermont-Ferrand, France\\
\email{yan.gerard@uca.fr} \and
LIS, Aix--Marseille Universit\'e, Marseille, France\\
\email{guilherme.fonseca@lis-lab.fr}}

\providecommand{\Example}{
\begin{tikzpicture}[
    x=0.012cm, y=-0.012cm,
    trait/.style={draw=blue, line width=0.8pt,
                  line cap=round, line join=round}, scale=0.87,
    pointille/.style={trait, dash pattern=on 2.2pt off 2.2pt}
]

\begin{scope}[shift={(-3,0)}]
\draw[trait] (7,53)--(52,76)--(42,112)--(93,119)--(115,54)
             --(135,118)--(71,172)--(3,136);
\foreach \p in {(7,53),(52,76),(42,112),(93,119),(115,54),(135,118),(71,172),(3,136)} {\fill[black] \p circle[radius=1.15pt];}
\end{scope}

\begin{scope}[shift={(3,50)}]
\draw[trait] (216,26)--(205,63)--(256,69)--(278,5)--(299,68)
             --(234,122)--(167,87)--(172,3);
\draw[pointille] (172,3)--(216,26);
\foreach \p in {(172,3),(216,26),(205,63),(256,69),(278,5),(299,68),(234,122),(167,87)} {\fill[black] \p circle[radius=1.15pt];}
\end{scope}

\begin{scope}[shift={(-10,43)}]
\draw[trait] (354,10)--(399,33);
\draw[pointille] (399,33)--(388,69);
\draw[trait] (388,69)--(439,77)--(462,12)--(482,75)
             --(417,129)--(350,94)--(354,10);
\foreach \p in {(354,10),(399,33),(388,69),(439,77),(462,12),(482,75),(417,129),(350,94)} {\fill[black] \p circle[radius=1.15pt];}
\end{scope}

\begin{scope}[shift={(-14,-30)}]
\draw[trait] (529,83)--(574,107)--(563,143);
\draw[pointille] (563,143)--(615,150);
\draw[trait] (615,150)--(637,85)--(657,148)
             --(592,202)--(524,167)--(529,83);
\foreach \p in {(529,83),(574,107),(563,143),(615,150),(637,85),(657,148),(592,202),(524,167)} {\fill[black] \p circle[radius=1.15pt];}
\end{scope}

\begin{scope}[shift={(669,-226)}]
\draw[trait] (11,364)--(15,279)--(49,340)--(60,303)
             --(78,398)--(102,347)--(143,345)--(123,282);
\draw[pointille] (102,347)--(123,282);
\foreach \p in {(11,364),(15,279),(49,340),(60,303),(78,398),(102,347),(143,345),(123,282)} {\fill[black] \p circle[radius=1.15pt];}
\end{scope}

\begin{scope}[shift={(669,-252)}]
\draw[trait] (181,390)--(185,305)--(219,366)--(230,330)
             --(248,425)--(270,373)--(293,308)--(313,371);
\draw[pointille] (313,371)--(248,425);
\foreach \p in {(181,390),(185,305),(219,366),(230,330),(248,425),(270,373),(293,308),(313,371)} {\fill[black] \p circle[radius=1.15pt];}
\end{scope}

\begin{scope}[shift={(667,-246)}]
\draw[trait] (353,382)--(358,299)--(392,359)--(403,322)
             --(421,419)--(487,365)--(466,302)--(444,366);
\draw[pointille] (353,382)--(421,419);
\foreach \p in {(353,382),(358,299),(392,359),(403,322),(421,419),(487,365),(466,302),(444,366)} {\fill[black] \p circle[radius=1.15pt];}
\end{scope}

\end{tikzpicture}
}

\providecommand{\Tree}{
\begin{tikzpicture}[
    every node/.style={ inner sep=3pt},
    line width=0.8pt,scale=0.6
]

\node (T)  at ( 0.0,  0.0) {Theorem 1};

\node (L2) at (-4.0, -1.0) {Lemma 2};
\node (L3) at (-1.8, -1.6) {Lemma 3};
\node (L4) at ( 1.4, -1.6) {Lemma 4};
\node (L5) at ( 4.1, -1.0) {Lemma 5};

\node (L6) at (-1.5, -3.0) {Lemma 6};
\node (L7) at ( 3.3, -3.0) {Lemma 7};
\node (L10) at (5.9, -2.6) {Lemma 10};

\node (L8) at ( 1.3, -4.6) {Lemma 8};
\node (L9) at ( 4.7, -5.8) {Lemma 9};

\draw (T) -- (L2);
\draw (T) -- (L3);
\draw (T) -- (L4);
\draw (T) -- (L5);

\draw (L4) -- (L6);
\draw (L4) -- (L7);

\draw (L5) -- (L10);

\draw (L6) -- (L8);
\draw (L7) -- (L8);
\draw (L7) -- (L9);
\draw (L8) -- (L9);

\end{tikzpicture}
}

\providecommand{\FigureEdgeTypes}{
\begin{tikzpicture}[scale=0.9]

\foreach \i in {0,1,2,3,4,5} {
    \coordinate (O\i) at (60*\i:2.5);
}
\draw[dashed] (O0)--(O1)--(O2)--(O3)--(O4)--(O5)--cycle;

\foreach \i in {0,1,2,3,4} {
    \coordinate (I\i) at (72*\i:1.5);
}
\draw[dashed] (I0)--(I1)--(I2)--(I3)--(I4)--cycle;
\fill[gray!10] (I0)--(I1)--(I2)--(I3)--(I4)--cycle;

\draw[blue,very thick] (I0)--(I1);
\node[blue] at ($(I0)!0.5!(I1)+(1,0)$) {\small level edge};

\draw[green!70!black,very thick] (O1)--(I1);
\node[green!70!black] at ($(O1)!0.7!(I1)+(0.8,0)$) {\small bridge};

\draw[orange,very thick] (I3)--(I0);
\node[orange] at ($(I3)!0.3!(I0)+(0,0.25)$) {\small small diagonal};

\draw[purple,very thick] (O3)--(I1);
\node[purple] at ($(O3)!0.5!(I1)+(-0.2,-0.2)$) {\small medium diagonal};

\draw[red,very thick] (O4)--(O0);
\node[red] at ($(O4)!0.2!(O0)+(1.4,0)$) {\small big diagonal};

\foreach \i in {0,1,2,3,4,5} {
    \fill (O\i) circle (2pt);
}
\foreach \i in {0,1,2,3,4} {
    \fill (I\i) circle (2pt);
}

\node at (0,2.5) {$L_0$};
\node at (-0.5,1.5) {$L_1$};

\end{tikzpicture}
}

\providecommand{\FigureLeastPotential}[1]{
\begin{tikzpicture}[scale=0.8]

\coordinate (p1) at (0.00,0.00);
\coordinate (p2) at (1.18,0.34);
\coordinate (p3) at (1.67,-0.17);
\coordinate (p4) at (1.21,-0.46);
\coordinate (p5) at (0.51,-0.50);
\coordinate (p6) at (0.86,-1.24);
\coordinate (p7) at (-0.38,-0.92);
\coordinate (p8) at (-0.07,0.59);
\coordinate (p9) at (2.56,0.45);
\coordinate (p10) at (2.13,-0.59);

\fill[gray!10] (p1)--(p2)--(p3)--(p4)--(p5)--cycle;

\draw[blue] (p5) -- (p1) -- (p2) -- (p3);
\draw[blue] (p6)-- (p7) -- (p8) -- (p9) -- (p10);
#1

\foreach \i in {1,...,10}
    \filldraw[black!70] (p\i) circle (1pt);

\fill[black] (p4) circle (1.5pt);
\fill[black] (p5) circle (1.5pt);
\fill[black] (p6) circle (1.5pt);
\node[above] at (p4) {\small $v'$};
\node[above] at (p5) {\small $v$};
\node[right] at (p6) {\small $u$};

\end{tikzpicture}
}
\providecommand{\FigureLeastPotentiali}{
    \FigureLeastPotential{
        \draw[black!70, dashed] (p4) -- (p6);
        \draw[blue] (p3) -- (p4);
        \draw[blue,thick] (p5)--(p6);
    }
}
\providecommand{\FigureLeastPotentialii}{
    \FigureLeastPotential{
        \draw[black!70, dashed] (p4) -- (p5);
        \draw[blue] (p3) -- (p4);
        \draw[blue,thick] (p4)--(p6);
    }
}
\providecommand{\FigureLeastPotentialiii}{
    \FigureLeastPotential{
        \draw[blue] (p4) -- (p5);
        \draw[blue,thick] (p4)--(p6);
    }
}

\providecommand{\FigureObstrCaseI}{
\begin{tikzpicture}[scale=0.8]
\coordinate (p1) at (0.00,0.00);
\coordinate (p2) at (-0.20,1.26);
\coordinate (p3) at (-0.96,0.82);
\coordinate (p4) at (-1.08,2.00);
\coordinate (p5) at (-1.38,-0.34);
\coordinate (p6) at (0.44,-0.80);
\coordinate (p7) at (0.91,1.45);
\coordinate (p8) at (-0.13,2.24);
\coordinate (p9) at (2.12,1.61);
\coordinate (p10) at (2.23,-0.70);
\coordinate (p11) at (1.36,-0.09);
\coordinate (p12) at (1.73,0.75);

\fill[gray!10] (p1)--(p3)--(p2)--(p7)--(p12)--(p11)--cycle;

\draw[blue] (p1) -- (p2) -- (p3) -- (p4) -- (p5) -- (p6);
\draw[blue] (p7) -- (p8) -- (p9) -- (p10) -- (p11) -- (p12);

\draw[thick, blue] (p6)--(p7);
\draw[thick, black!70, dashed] (p1)--(p7);

\foreach \i in {1,...,12}
    \filldraw[black!70] (p\i) circle (1pt);

\foreach \i/\dir/\name in {
    1/below/$s$,
    6/right/$u$,
    7/right/$v$,
    11/below/$q$
}{
    \fill[black] (p\i) circle (1.5pt);
    \node[\dir] at (p\i) {\small \name};
}

\end{tikzpicture}
}

\providecommand{\FigureObstrCaseII}{
\begin{tikzpicture}[scale=0.8]
\coordinate (p1) at (0.00,0.00);
\coordinate (p2) at (-0.84,0.51);
\coordinate (p3) at (0.65,1.34);
\coordinate (p4) at (0.60,-1.04);
\coordinate (p5) at (-1.53,-0.64);
\coordinate (p6) at (-1.20,1.90);
\coordinate (p7) at (-0.39,1.23);
\coordinate (p8) at (1.56,1.91);
\coordinate (p9) at (1.14,0.02);
\coordinate (p10) at (1.99,-0.52);

\fill[gray!10] (p1)--(p2)--(p7)--(p3)--(p9)--cycle;

\draw[blue] (p1) -- (p2) -- (p3); 
\draw[blue] (p4) -- (p5) -- (p6) -- (p7) -- (p8) -- (p9) -- (p10);

\draw[thick, blue] (p3)--(p4);
\draw[thick, black!70, dashed] (p1)--(p4);

\foreach \i in {1,...,10}
    \filldraw[black!70] (p\i) circle (1pt);
    
\foreach \i/\dir/\name in {
    1/below/$s$,
    3/right/$u$,
    4/right/$v$,
    9/below/$q$
}{
    \fill[black] (p\i) circle (1.5pt);
    \node[\dir] at (p\i) {\small \name};
}
\end{tikzpicture}
}

\providecommand{\FigureObstrCaseIII}{
\begin{tikzpicture}[scale=0.8]
\coordinate (p1) at (0.00,0.00);
\coordinate (p2) at (-0.12,1.52);
\coordinate (p3) at (-0.91,0.42);
\coordinate (p4) at (-0.73,2.18);
\coordinate (p5) at (-1.81,0.79);
\coordinate (p6) at (-0.02,-0.78);
\coordinate (p7) at (1.26,2.26);
\coordinate (p8) at (0.78,0.04);
\coordinate (p9) at (1.37,0.68);
\coordinate (p10) at (1.66,0.06);

\fill[gray!10] (p1)--(p3)--(p2)--(p9)--(p8);

\draw[blue] (p1) -- (p2) -- (p3) -- (p4) -- (p5) -- (p6);
\draw[blue] (p7) -- (p8) -- (p9) -- (p10);

\draw[thick, blue] (p6)--(p7);
\draw[thick, black!70, dashed] (p1)--(p7);

\foreach \i in {1,...,10}
    \filldraw[black!70] (p\i) circle (1pt);

\foreach \i/\dir/\name in {
    1/below/$s$,
    6/right/$u$,
    7/right/$v$,
    8/below/$q$
}{
    \fill[black] (p\i) circle (1.5pt);
    \node[\dir] at (p\i) {\small \name};
}
\end{tikzpicture}
}

\providecommand{\FigureObstrCaseIV}[1]{
\begin{tikzpicture}[scale = 0.8]

\coordinate (p1) at (0.00,0.00);
\coordinate (p2) at (-0.60,0.23);
\coordinate (p3) at (-1.21,1.84);
\coordinate (p4) at (-0.11,2.20);
\coordinate (p5) at (-1.18,3.20);
\coordinate (p6) at (-0.94,2.11);
\coordinate (p7) at (-2.08,2.34);
\coordinate (p8) at (-1.30,0.83);
\coordinate (p9) at (-1.32,1.48);
\coordinate (p10) at (-0.99,0.45);
\coordinate (p11) at (-2.14,1.04);
\coordinate (p12) at (-1.33,-0.67);
\coordinate (p13) at (1.72,-1.06);
\coordinate (p14) at (-0.56,3.35);
\coordinate (p15) at (1.92,2.67);
\coordinate (p16) at (2.97,0.50);
\coordinate (p17) at (1.36,0.69);
\coordinate (p18) at (0.90,1.88);

\filldraw[gray!10] (p1)--(p2)--(p10)--(p8)--(p9)--(p3)--(p6)--(p4)--(p18)--(p17)--cycle;

#1
\draw[thick, blue] (p4) -- (p5) -- (p6) -- (p7) -- (p8) -- (p9) -- (p10) -- (p11) -- (p12) -- (p13);
\draw[thick, red] (p13) -- (p14);
\draw[blue] (p14) -- (p15) -- (p16) -- (p17) -- (p18);

\foreach \i in {1,...,18}
    \filldraw[black!70] (p\i) circle (1pt);

\foreach \i/\dir/\name in {
    1/below/$s$,
    4/above right/$p$,
    5/left/$v'$,
    6/above left/$p'$,
    13/right/$u$,
    14/left/$v$,
    17/below/$q$
}{
    \fill[black] (p\i) circle (1.5pt);
    \node[\dir] at (p\i) {\small \name};
}

\end{tikzpicture}
}
\providecommand{\FigurePrefixStructure}{
    \FigureObstrCaseIV{
        \filldraw[dotted, red!30, opacity=0.4] (p13) -- (p1) -- (p14);
        \draw[thick, violet] (p1) -- (p2) -- (p3) -- (p4);
    }
}

\providecommand{\FigureObstrCaseIVdI}[1]{
\begin{tikzpicture}[scale=0.8]
\coordinate (p1) at (-0.25,0.00);
\coordinate (p2) at (-0.82,0.08);
\coordinate (p3) at (0.27,2.30);
\coordinate (p4) at (-1.10,1.72);
\coordinate (p5) at (-0.71,2.17);
\coordinate (p6) at (-1.73,2.29);
\coordinate (p7) at (-1.20,0.67);
\coordinate (p8) at (-1.66,-0.25);
\coordinate (p9) at (0.75,-0.40);
\coordinate (p10) at (0.32,2.90);
\coordinate (p11) at (0.98,1.54);
\coordinate (p12) at (1.68,1.30);
\coordinate (p13) at (0.93,0.29);
\coordinate (p14) at (1.21,0.87);

\filldraw[gray!10] (p1)--(p2)--(p7)--(p4)--(p5)--(p3)--(p11)--(p14)--(p13)--cycle;

\draw[blue, thick] (p1) -- (p2);
\draw[blue, thick] (p4) -- (p5) -- (p6) -- (p7) -- (p8) -- (p9);
\draw[blue] (p10) -- (p11) -- (p12) -- (p13) -- (p14);
#1

\foreach \i in {1,...,14}
    \filldraw[black!70] (p\i) circle (1pt);

\foreach \i/\dir/\name in {
    1/below/$s$,
    2/left/$s'$,
    3/above left/$p$,
    9/right/$u$,
    10/right/$v$,
    13/below/$q$
}{
    \fill[black] (p\i) circle (1.5pt);
    \node[\dir] at (p\i) {\small \name};
}
\end{tikzpicture}
}
\providecommand{\FigureObstrCaseIVdIi}{
    \FigureObstrCaseIVdI{
        \draw[black!70, dashed, thick] (p1)--(p3);
        \draw[thick, blue] (p2)--(p3)--(p4);
        \draw[very thick, red] (p9)--(p10);
    }
}
\providecommand{\FigureObstrCaseIVdIii}{
    \FigureObstrCaseIVdI{
        \draw[black!70, dashed, thick] (p2)--(p4);
        \draw[thick, blue] (p1)--(p3)--(p4);
        \draw[very thick, red] (p9)--(p10);
    }
}
\providecommand{\FigureObstrCaseIVdIiii}{
    \FigureObstrCaseIVdI{
        \draw[black!70, dashed, thick] (p3)--(p10);
        \draw[thick, blue] (p4)--(p2)--(p1)--(p3);
        \draw[very thick, red] (p9)--(p10);
    }
}
\providecommand{\FigureObstrCaseIVdIiv}{
    \FigureObstrCaseIVdI{
        \draw[black!70, dashed, thick] (p3)--(p10);
        \draw[thick, blue] (p4)--(p2)--(p1)--(p3)--(p10);
    }
}

\providecommand{\FigureObstrCaseIVdIIdI}[1]{
\begin{tikzpicture}[scale=0.7]
\coordinate (p1) at (-0.50,0.00);
\coordinate (p2) at (-1.49,0.46);
\coordinate (p3) at (-0.93,1.91);
\coordinate (p4) at (0.35,2.15);
\coordinate (p5) at (-0.71,2.78);
\coordinate (p6) at (-1.47,1.07);
\coordinate (p7) at (-1.76,2.37);
\coordinate (p8) at (-2.36,0.96);
\coordinate (p9) at (-1.67,-0.67);
\coordinate (p10) at (-1.00,-0.03);
\coordinate (p11) at (0.37,-0.92);
\coordinate (p12) at (0.50,2.85);
\coordinate (p13) at (1.51,2.27);
\coordinate (p14) at (1.14,0.65);
\coordinate (p15) at (1.54,0.08);

\filldraw[gray!10] (p1)--(p10)--(p2)--(p6)--(p3)--(p4)--(p14)--cycle;

\draw[thick, blue] (p1) -- (p2) -- (p3);
\draw[thick, blue] (p5) -- (p6) -- (p7) -- (p8) -- (p9) -- (p10) -- (p11);
\draw[blue] (p12) -- (p13) -- (p14) -- (p15);
#1
\foreach \i in {1,...,15}
    \filldraw[black!70] (p\i) circle (1pt);

\foreach \i/\dir/\name in {
    1/below right/$s$,
    3/below right/$s'$,
    4/above left/$p$,
    5/left/$v'$,
    11/right/$u$,
    12/right/$v$,
    14/below/$q$
}{
    \fill[black] (p\i) circle (1.5pt);
    \node[\dir] at (p\i) {\small \name};
}
\end{tikzpicture}
}
\providecommand{\FigureObstrCaseIVdIIdIi}{
    \FigureObstrCaseIVdIIdI{
        \draw[thick, black!70, dashed] (p1)--(p4);
        \draw[thick, blue] (p3)--(p4)--(p5);
        \draw[very thick, red] (p11)--(p12);
    }
}
\providecommand{\FigureObstrCaseIVdIIdIii}{
    \FigureObstrCaseIVdIIdI{
        \draw[thick, black!70, dashed] (p3)--(p5);
        \draw[thick, blue] (p1)--(p4)--(p5);
        \draw[very thick, red] (p11)--(p12);
    }
}
\providecommand{\FigureObstrCaseIVdIIdIiii}{
    \FigureObstrCaseIVdIIdI{
        \draw[thick, black!70, dashed] (p4)--(p12);
        \draw[thick, blue] (p1)--(p4) (p3)--(p5);
        \draw[very thick, red] (p11)--(p12);
    }
}
\providecommand{\FigureObstrCaseIVdIIdIiv}{
    \FigureObstrCaseIVdIIdI{
        \draw[thick, blue] (p1)--(p4)--(p12) (p3)--(p5);
    }
}

\providecommand{\FigureObstrCaseIVdIIdII}[1]{
\begin{tikzpicture}[scale = 0.8]
\coordinate (p1) at (0.00,0.00);
\coordinate (p2) at (-0.56,0.11);
\coordinate (p3) at (0.36,2.65);
\coordinate (p4) at (-0.40,3.34);
\coordinate (p5) at (-1.28,3.10);
\coordinate (p6) at (-1.48,0.76);
\coordinate (p7) at (-1.84,1.54);
\coordinate (p8) at (-1.59,2.54);
\coordinate (p9) at (-2.43,2.03);
\coordinate (p10) at (-2.03,0.59);
\coordinate (p11) at (-1.12,0.37);
\coordinate (p12) at (-0.42,-0.40);
\coordinate (p13) at (1.26,-0.40);
\coordinate (p14) at (0.27,3.39);
\coordinate (p15) at (1.53,2.76);
\coordinate (p16) at (1.21,2.32);
\coordinate (p17) at (1.24,0.37);
\coordinate (p18) at (1.65,1.42);
\coordinate (p19) at (1.99,0.62);

\filldraw[gray!10] (p1)--(p2)--(p11)--(p6)--(p7)--(p8)--(p3)--(p16)--(p18)--(p17)--cycle;

\draw[thick, blue] (p1) -- (p2) (p3) -- (p4) -- (p5) (p6) -- (p7) -- (p8) -- (p9) -- (p10) -- (p11) -- (p12) -- (p13);
\draw[blue] (p14) -- (p15) -- (p16) -- (p17) -- (p18) -- (p19);
\draw[thick, red] (p13)--(p14);
#1

\foreach \i in {1,...,19}
    \filldraw[black!70] (p\i) circle (1pt);

\foreach \i/\dir/\name in {
    1/right/$s$,
    2/above/$s'$,
    3/right/$p$,
    5/left/$a$,
    6/left/$b$,
    8/left/$p'$
}{
    \fill[black] (p\i) circle (1.5pt);
    \node[\dir] at (p\i) {\small \name};
}
\end{tikzpicture}

}
\providecommand{\FigureObstrCaseIVdIIdIIi}{
    \FigureObstrCaseIVdIIdII{
        \draw[thick, black!70, dashed] (p1)--(p3);
        \draw[thick, blue] (p2)--(p3);
        \draw[thick, violet] (p5)--(p6);
    }
}
\providecommand{\FigureObstrCaseIVdIIdIIii}{
    \FigureObstrCaseIVdIIdII{
        \draw[thick, black!70, dashed] (p2)--(p6);
        \draw[thick, blue] (p1)--(p3);
        \draw[thick, violet] (p5)--(p6);
    }
}
\providecommand{\FigureObstrCaseIVdIIdIIiii}{
    \FigureObstrCaseIVdIIdII{
        \draw[thick, blue] (p2)--(p6);
        \draw[thick, blue] (p1)--(p3);
    }
}

\providecommand{\FigureObstrCaseIVdIIdIIIi}{
    \FigureObstrCaseIV{
        \draw[thick, black!70, dashed] (p4)--(p1);
        \draw[thick, blue] (p1)--(p2)--(p3)--(p4);
        \fill[black] (p3) circle (1.5pt);
        \node[left] at (p3) {\small $s'$};
    }
}
\providecommand{\FigureObstrCaseIVdIIdIIIii}{
    \FigureObstrCaseIV{
        \draw[thick, blue] (p4)--(p1)--(p2)--(p3);
        \fill[black] (p3) circle (1.5pt);
        \node[left] at (p3) {\small $s'$};
    }
}

\providecommand{\FigureObstrCaseIVdIIdIIIdI}[1]{
\begin{tikzpicture}[scale=0.7]
\coordinate (p1) at (0.00,0.00);
\coordinate (p2) at (0.60,-0.23);
\coordinate (p3) at (0.49,1.97);
\coordinate (p4) at (-0.58,2.97);
\coordinate (p5) at (-0.34,1.88);
\coordinate (p6) at (-0.70,1.60);
\coordinate (p7) at (-1.48,2.11);
\coordinate (p8) at (-0.70,0.60);
\coordinate (p9) at (-0.82,1.05);
\coordinate (p10) at (-0.39,0.22);
\coordinate (p11) at (-1.34,0.71);
\coordinate (p12) at (-0.73,-0.60);
\coordinate (p13) at (2.02,-0.69);
\coordinate (p14) at (0.04,3.12);
\coordinate (p15) at (1.50,1.65);
\coordinate (p16) at (2.22,2.24);
\coordinate (p17) at (1.96,0.46);

\filldraw[gray!10] (p2)--(p1)--(p10)--(p8)--(p9)--(p6)--(p5)--(p3)--(p15)--(p17)--cycle;

\draw[thick, blue] (p1)--(p2)--(p3) (p4)--(p5) (p6) -- (p7) -- (p8) -- (p9) -- (p10) -- (p11) -- (p12) -- (p13);
\draw[blue] (p14) -- (p15) -- (p16) -- (p17);
#1

\foreach \i in {1,...,17}
    \filldraw[black!70] (p\i) circle (1pt);

\foreach \i/\dir/\name in {
    1/below/$s'$,
    2/below/$s$,
    3/above right/$p$,
    4/left/$v'$,
    5/above left/$p'$,
    13/right/$u$,
    14/right/$v$
}{
    \fill[black] (p\i) circle (1.5pt);
    \node[\dir] at (p\i) {\small \name};
}
\end{tikzpicture}
}
\providecommand{\FigureObstrCaseIVdIIdIIIdIi}{
    \FigureObstrCaseIVdIIdIIIdI{
        \draw[thick, black!70, dashed] (p1)--(p6);
        \draw[thick, blue] (p3)--(p4) (p5)--(p6);
        \draw[very thick, red] (p13)--(p14);
    }
}
\providecommand{\FigureObstrCaseIVdIIdIIIdIii}{
    \FigureObstrCaseIVdIIdIIIdI{
        \draw[thick, black!70, dashed] (p3)--(p5);
        \draw[thick, blue] (p3)--(p4) (p1)--(p6);
        \draw[very thick, red] (p13)--(p14);
    }
}
\providecommand{\FigureObstrCaseIVdIIdIIIdIiii}{
    \FigureObstrCaseIVdIIdIIIdI{
        \draw[thick, black!70, dashed] (p4)--(p14);
        \draw[thick, blue] (p3)--(p5) (p1)--(p6);
        \draw[very thick, red] (p13)--(p14);
    }
}
\providecommand{\FigureObstrCaseIVdIIdIIIdIiv}{
    \FigureObstrCaseIVdIIdIIIdI{
        \draw[thick, blue] (p3)--(p5) (p1)--(p6) (p4)--(p14);
    }
}

\providecommand{\FigureObstrCaseIVdIIdIIIdII}[1]{
\begin{tikzpicture}[scale=0.8]
\coordinate (p1) at (0.00,0.00);
\coordinate (p2) at (0.78,0.08);
\coordinate (p3) at (0.62,2.28);
\coordinate (p4) at (-0.40,3.28);
\coordinate (p5) at (-1.22,2.77);
\coordinate (p6) at (-0.55,2.15);
\coordinate (p7) at (-1.76,1.51);
\coordinate (p8) at (-1.21,1.30);
\coordinate (p9) at (-1.74,0.21);
\coordinate (p10) at (-0.97,0.59);
\coordinate (p11) at (-0.52,0.13);
\coordinate (p12) at (-0.73,-0.60);
\coordinate (p13) at (2.02,-0.49);
\coordinate (p14) at (0.22,3.43);
\coordinate (p15) at (2.42,2.24);
\coordinate (p16) at (2.14,0.77);
\coordinate (p17) at (1.68,1.96);

\filldraw[gray!10] (p2)--(p1)--(p11)--(p10)--(p8)--(p6)--(p3)--(p17)--(p16)--cycle;

\draw[thick, blue] (p1)--(p2)--(p3)--(p4)--(p5) (p7) -- (p8) -- (p9) -- (p10) -- (p11) -- (p12) -- (p13);
\draw[blue] (p14) -- (p15) -- (p16) -- (p17);
\draw[thick, red] (p13)--(p14);
#1

\foreach \i in {1,...,17}
    \filldraw[black!70] (p\i) circle (1pt);

\foreach \i/\dir/\name in {
    1/below/$s'$,
    2/below/$s$,
    3/above right/$p$,
    5/above/$x$,
    6/above right/$p'$,
    7/above/$y$
}{
    \fill[black] (p\i) circle (1.5pt);
    \node[\dir] at (p\i) {\small \name};
}

\end{tikzpicture}
}

\providecommand{\FigureObstrCaseIVdIIdIIIdIIi}{
    \FigureObstrCaseIVdIIdIIIdII{
        \draw[black!70, thick, dashed] (p1)--(p6);
        \draw[violet, thick] (p5)--(p6)--(p7);
    }
}
\providecommand{\FigureObstrCaseIVdIIdIIIdIIii}{
    \FigureObstrCaseIVdIIdIIIdII{
        \draw[black!70, thick, dashed] (p5)--(p7);
        \draw[violet, thick] (p6)--(p7);
        \draw[blue, thick] (p1)--(p6);
    }
}
\providecommand{\FigureObstrCaseIVdIIdIIIdIIiii}{
    \FigureObstrCaseIVdIIdIIIdII{
        \draw[blue, thick] (p1)--(p6) (p5)--(p7);
    }
}

\providecommand{\FigureObstrCaseIVdIIdIIIdIIIdIII}[1]{
\begin{tikzpicture}[scale=0.8]
\coordinate (p1) at (0.00,0.00);
\coordinate (p2) at (0.78,0.08);
\coordinate (p3) at (0.67,2.28);
\coordinate (p4) at (-0.40,3.28);
\coordinate (p5) at (-0.87,1.86);
\coordinate (p6) at (-1.76,1.51);
\coordinate (p7) at (-1.21,1.30);
\coordinate (p8) at (-0.16,2.19);
\coordinate (p9) at (-1.74,0.21);
\coordinate (p10) at (-0.97,0.59);
\coordinate (p11) at (-0.52,0.13);
\coordinate (p12) at (-0.73,-0.60);
\coordinate (p13) at (2.22,-0.29);
\coordinate (p14) at (0.2,3.43);
\coordinate (p15) at (2.42,2.24);
\coordinate (p16) at (1.50,1.65);
\coordinate (p17) at (1.96,0.46);

\filldraw[gray!10] (p2)--(p1)--(p11)--(p10)--(p7)--(p5)--(p8)--(p3)--(p16)--(p17)--cycle;

\draw[thick, blue] (p1) -- (p2) -- (p3) -- (p4) -- (p5) -- (p6) -- (p7) -- (p8)  (p9) -- (p10) (p11) -- (p12) -- (p13);
\draw[thick, red] (p13) -- (p14);
\draw[gray] (p14) -- (p15) -- (p16) -- (p17);
#1

\foreach \i in {1,...,17}
    \filldraw[black!70] (p\i) circle (1pt);

\foreach \i/\dir/\name in {
    1/below/$s'$,
    2/below/$s$,
    3/above right/$p$,
    8/above/$p'$
}{
    \fill[black] (p\i) circle (1.5pt);
    \node[\dir] at (p\i) {\small \name};
}
\end{tikzpicture}
}
\providecommand{\FigureObstrCaseIVdIIdIIIdIIIdIIIi}{
    \FigureObstrCaseIVdIIdIIIdIIIdIII{
        \draw[thick, black!70, dashed] (p1)--(p11);
        \draw[thick, violet] (p8)--(p9);
        \draw[thick, blue] (p10)--(p11);
    }
}
\providecommand{\FigureObstrCaseIVdIIdIIIdIIIdIIIii}{
    \FigureObstrCaseIVdIIdIIIdIIIdIII{
        \draw[thick, black!70, dashed] (p8)--(p10);
        \draw[thick, violet] (p8)--(p9);
        \draw[thick, blue] (p1)--(p11);
    }
}
\providecommand{\FigureObstrCaseIVdIIdIIIdIIIdIIIiii}{
    \FigureObstrCaseIVdIIdIIIdIIIdIII{
        \draw[thick, blue] (p1)--(p11) (p8)--(p10);
    }
}

\providecommand{\FigureEndpointOutPrefixStructure}{
\begin{tikzpicture}[scale=0.8]
\coordinate (p1) at (0.00,0.00);
\coordinate (p2) at (-1.15,0.26);
\coordinate (p3) at (-2.06,1.31);
\coordinate (p4) at (-1.24,1.35);
\coordinate (p5) at (-0.54,0.57);
\coordinate (p6) at (-0.97,2.54);
\coordinate (p7) at (-1.36,2.04);
\coordinate (p8) at (-1.46,3.34);
\coordinate (p9) at (0.31,3.56);
\coordinate (p10) at (0.56,2.50);
\coordinate (p11) at (1.67,2.68);
\coordinate (p12) at (1.84,0.68);
\coordinate (p13) at (0.94,0.13);
\coordinate (p14) at (0.20,0.74);
\coordinate (p15) at (-0.25,2.77);
\coordinate (p16) at (0.93,1.29);
\coordinate (p17) at (0.89,1.80);

\filldraw[gray!10] (p5)--(p4)--(p7)--(p6)--(p15)--(p10)--(p17)--(p16)--(p14)--cycle;
\draw[thick, blue] (p1)--(p2) -- (p3) -- (p4) -- (p5) -- (p6) -- (p7) -- (p8) -- (p9) -- (p10) -- (p11) -- (p12)--(p13);
\draw[very thick, red] (p13) -- (p14); 
\draw[thick, violet] (p14)-- (p15) -- (p16) -- (p17);

\draw[black!70, dashed] (p1)--(p2)--(p3)--(p8)--(p9)--(p11)--(p12)--(p13)--cycle;

\foreach \i in {1,...,17}
    \filldraw[black!70] (p\i) circle (1pt);

\foreach \i/\dir/\name in {
    1/above/$s$,
    13/above/$s'$,
    17/right/$t$
}{
    \fill[black] (p\i) circle (1.5pt);
    \node[\dir] at (p\i) {\small \name};
}

\end{tikzpicture}
}

\providecommand{\FigureEndpointOutCaseIIIdIIIdI}[1]{
\begin{tikzpicture}[scale=0.8]
\coordinate (p1) at (0.00,0.00);
\coordinate (p2) at (-1.15,0.26);
\coordinate (p3) at (-2.06,1.31);
\coordinate (p4) at (-1.24,1.35);
\coordinate (p5) at (-0.54,0.57);
\coordinate (p6) at (-0.97,2.54);
\coordinate (p7) at (-1.36,2.04);
\coordinate (p8) at (-1.46,3.34);
\coordinate (p9) at (0.31,3.56);
\coordinate (p10) at (-0.32,2.84);
\coordinate (p11) at (0.83,2.33);
\coordinate (p12) at (0.45,2.78);
\coordinate (p13) at (1.77,2.38);
\coordinate (p14) at (0.94,0.13);
\coordinate (p15) at (0.20,0.74);
\coordinate (p16) at (0.93,1.29);

\filldraw[gray!10] (p5)--(p4)--(p7)--(p6)--(p10)--(p12)--(p11)--(p16)--(p15)--cycle;

\draw[thick, blue] (p1) -- (p2) -- (p3) -- (p4) -- (p5) -- (p6) -- (p7) -- (p8) -- (p9) -- (p10) -- (p11)  (p12) -- (p13) -- (p14) (p15) -- (p16);
#1

\foreach \i in {1,...,16}
    \filldraw[black!70] (p\i) circle (1pt);

\foreach \i/\dir/\name in {
    1/above/$s$,
    14/above/$s'$,
    11/right/$t'$,
    16/left/$t$
}{
    \fill[black] (p\i) circle (1.5pt);
    \node[\dir] at (p\i) {\small \name};
}
\end{tikzpicture}
}
\providecommand{\FigureEndpointOutCaseIIIdIIIdIi}{
    \FigureEndpointOutCaseIIIdIIIdI{
        \draw[thick, black!70, dashed] (p11)--(p16);
        \draw[thick, blue] (p11)--(p12);
        \draw[very thick, red] (p14)--(p15);
    }
}
\providecommand{\FigureEndpointOutCaseIIIdIIIdIii}{
    \FigureEndpointOutCaseIIIdIIIdI{
        \draw[thick, black!70, dashed] (p1)--(p14);
        \draw[thick, blue] (p11)--(p16);
        \draw[very thick, red] (p14)--(p15);
    }
}
\providecommand{\FigureEndpointOutCaseIIIdIIIdIiii}{
    \FigureEndpointOutCaseIIIdIIIdI{
        \draw[thick, blue] (p11)--(p16) (p1)--(p14);
    }
}

\providecommand{\FigureEndpointOutCaseIIIdIIIdII}[1]{
\begin{tikzpicture}[scale=0.8]
\coordinate (p1) at (0.00,0.00);
\coordinate (p2) at (-1.15,0.26);
\coordinate (p3) at (-2.06,1.31);
\coordinate (p4) at (-1.24,1.35);
\coordinate (p5) at (-1.36,2.04);
\coordinate (p6) at (-1.74,3.04);
\coordinate (p7) at (0.31,3.56);
\coordinate (p8) at (0.45,2.78);
\coordinate (p9) at (1.87,2.28);
\coordinate (p10) at (0.94,0.13);
\coordinate (p11) at (0.20,0.74);
\coordinate (p12) at (-0.54,0.57);
\coordinate (p13) at (-0.97,2.54);
\coordinate (p14) at (-0.32,2.84);
\coordinate (p15) at (0.83,2.33);
\coordinate (p16) at (0.89,1.80);

\filldraw[gray!10] (p4)--(p5)--(p13)--(p14)--(p8)--(p15)--(p16)--(p11)--(p12)--cycle;

\draw[thick, blue] (p1) -- (p2) -- (p3) -- (p4) -- (p5)--(p6) -- (p7) -- (p8)--(p9) -- (p10) (p11) -- (p12) -- (p13) -- (p14) (p15) -- (p16);
\draw[thick, red] (p10) -- (p11);
#1

\foreach \i in {1,...,16}
    \filldraw[black!70] (p\i) circle (1pt);

\foreach \i/\dir/\name in {
    1/above/$s$,
    10/above/$s'$,
    8/above right/$v'$,
    14/above/$u$,
    15/right/$v$,
    11/above/$t'$,
    16/below/$t$
}{
    \fill[black] (p\i) circle (1.5pt);
    \node[\dir] at (p\i) {\small \name};
}
\end{tikzpicture}
}
\providecommand{\FigureEndpointOutCaseIIIdIIIdIIi}{
    \FigureEndpointOutCaseIIIdIIIdII{
        \draw[thick, black!70, dashed] (p14)--(p16);
        \draw[thick, violet] (p14)--(p15);
    }
}
\providecommand{\FigureEndpointOutCaseIIIdIIIdIIii}{
    \FigureEndpointOutCaseIIIdIIIdII{
        \draw[thick, blue] (p14)--(p16);
    }
}

\providecommand{\FigureEndpointsInCaseIIIdII}[3]{
\begin{tikzpicture}[scale=0.6]
\coordinate (p1) at (0.00,0.00);
\coordinate (p2) at (-0.99,0.13);
\coordinate (p3) at (-1.67,0.90);
\coordinate (p4) at (-0.80,2.75);
\coordinate (p5) at (-2.25,2.73);
\coordinate (p6) at (-1.65,1.97);
\coordinate (p7) at (-2.25,1.20);
\coordinate (p8) at (-1.03,-0.70);
\coordinate (p9) at (1.64,-0.51);
\coordinate (p10) at (1.58,0.57);
\coordinate (p11) at (0.73,0.10);
\coordinate (p12) at (0.20,2.90);
\coordinate (p13) at (-0.70,3.36);
\coordinate (p14) at (1.74,3.21);
\coordinate (p15) at (1.02,2.72);
\coordinate (p16) at (2.20,0.85);
\coordinate (p17) at (1.56,1.23);

\filldraw[gray!10] (p1)--(p2)--(p3)--(p6)--(p4)--(p12)--(p15)--(p17)--(p10)--(p11)--cycle;
\draw[blue] (p4) -- (p5) -- (p6)-- (p7) -- (p8) -- (p9) -- (p10) -- (p11);
\draw[thick, blue] (p10) -- (p11);
\draw[blue] (p12) -- (p13) -- (p14) -- (p15) -- (p16) -- (p17);
#1

\foreach \i in {1,...,17}
    \filldraw[black!70] (p\i) circle (1pt);

\foreach \i/\dir/\name in {
    #2/$s$,
    #3/$s'$,
    12/right/$x$
}{
    \fill[black] (p\i) circle (1.5pt);
    \node[\dir] at (p\i) {\small \name};
}
\end{tikzpicture}
}
\providecommand{\FigureEndpointsInCaseIIIdIIi}{
    \FigureEndpointsInCaseIIIdII{
        \draw[thick, black!70, dashed] (p12)--(p1)--(p11);
        \draw[thick, violet] (p11)--(p12);
        \draw[very thick, red] (p3)--(p4);
        \draw[thick, blue] (p1) -- (p2) -- (p3);
    }{1/below}{11/below}
}
\providecommand{\FigureEndpointsInCaseIIIdIIii}{
    \FigureEndpointsInCaseIIIdII{
        \draw[thick, black!70, dashed] (p12)--(p2)--(p1);
        \draw[thick, violet] (p1)--(p12);
        \draw[very thick, red] (p3)--(p4);
        \draw[thick, blue] (p11)-- (p1) (p2) -- (p3);
    }{2/below}{1/below}
}
\providecommand{\FigureEndpointsInCaseIIIdIIiii}{
    \FigureEndpointsInCaseIIIdII{
        \draw[thick, black!70, dashed] (p12)--(p3)--(p2);
        \draw[thick, violet] (p2)--(p12);
        \draw[very thick, red] (p3)--(p4);
        \draw[thick, blue] (p11)-- (p1)--(p2);
    }{3/below}{2/below}
}
\providecommand{\FigureEndpointsInCaseIIIdIIiv}{
    \FigureEndpointsInCaseIIIdII{
        \draw[thick, violet] (p3)--(p12);
        \draw[thick, blue] (p11)-- (p1)--(p2)--(p3);
    }{4/above left}{3/below}
}

\providecommand{\FigureEndpointsInCaseIIIdIIIdI}{
\begin{tikzpicture}[scale=0.7]
\coordinate (p1) at (0.00,0.00);
\coordinate (p2) at (-0.84,0.43);
\coordinate (p3) at (-1.09,2.13);
\coordinate (p4) at (0.52,2.58);
\coordinate (p5) at (-1.47,2.93);
\coordinate (p6) at (-1.91,1.67);
\coordinate (p7) at (-1.33,1.23);
\coordinate (p8) at (-1.45,-0.17);
\coordinate (p9) at (0.41,-0.50);
\coordinate (p10) at (2.02,-0.38);
\coordinate (p11) at (1.88,0.57);
\coordinate (p12) at (1.29,0.31);
\coordinate (p13) at (0.46,3.44);
\coordinate (p14) at (1.54,3.32);
\coordinate (p15) at (2.93,1.41);
\coordinate (p16) at (1.81,2.21);
\coordinate (p17) at (2.88,0.39);
\coordinate (p18) at (1.95,1.30);

\filldraw[gray!10] (p1)--(p2)--(p7)--(p3)--(p4)--(p16)--(p18)--(p11)--(p12)--cycle;

\filldraw[dotted, red!30, opacity=0.4] (p1)--(p12)--(p13)--cycle;

\draw[thick, violet] (p1) -- (p2) -- (p3) -- (p4);
\draw[thick, blue]  (p4) -- (p5) -- (p6) -- (p7) -- (p8) -- (p9) -- (p10);
\draw[blue] (p10) -- (p11) -- (p12) (p13) -- (p14) -- (p15) -- (p16) -- (p17) -- (p18);
\draw[very thick, red] (p12)--(p13);

\foreach \i in {1,...,18}
    \filldraw[black!70] (p\i) circle (1pt);

\foreach \i/\dir/\name in {
    1/below/$s$,
    4/below/$p$,
    10/right/$u'$,
    5/above right/$v'$,
    12/below/$u$,
    13/left/$v$
}{
    \fill[black] (p\i) circle (1.5pt);
    \node[\dir] at (p\i) {\small \name};
}
\end{tikzpicture}
}

\providecommand{\FigureEndpointsInCaseIIIdIIIdII}[3]{
\begin{tikzpicture}[scale=0.7]
\coordinate (p1) at (0.00,0.00);
\coordinate (p2) at (-0.86,-0.17);
\coordinate (p3) at (-2.13,0.12);
\coordinate (p4) at (-2.38,1.82);
\coordinate (p5) at (-1.47,2.23);
\coordinate (p6) at (-2.76,2.62);
\coordinate (p7) at (-3.20,1.36);
\coordinate (p8) at (-2.62,0.92);
\coordinate (p9) at (-2.74,-0.48);
\coordinate (p10) at (-0.88,-0.81);
\coordinate (p11) at (0.73,-0.69);
\coordinate (p12) at (1.59,0.08);
\coordinate (p13) at (0.66,0.99);
\coordinate (p14) at (0.59,0.26);
\coordinate (p15) at (-1.41,3.08);
\coordinate (p16) at (0.42,2.88);
\coordinate (p17) at (1.75,1.28);
\coordinate (p18) at (0.65,1.92);

\filldraw[gray!10] (p1)--(p2)--(p3)--(p8)--(p4)--(p5)--(p18)--(p13)--(p14)--cycle;

\draw[blue] (p4) -- (p5) -- (p6) -- (p7) -- (p8) -- (p9) -- (p10) -- (p11) -- (p12) -- (p13) -- (p14) (p15) -- (p16) -- (p17) -- (p18);
#1

\foreach \i in {1,...,18}
    \filldraw[black!70] (p\i) circle (1pt);

\foreach \i/\dir/\name in {
    #2/$s$,
    #3/$s'$,
    15/left/$x$
}{
    \fill[black] (p\i) circle (1.5pt);
    \node[\dir] at (p\i) {\small \name};
}
\end{tikzpicture}
}
\providecommand{\FigureEndpointsInCaseIIIdIIIdIIi}{
    \FigureEndpointsInCaseIIIdIIIdII{
        \draw[thick, black!70, dashed] (p15)--(p1)--(p14);
        \draw[thick, violet] (p14)--(p15);
        \draw[thick, blue] (p1) -- (p2) -- (p3)-- (p4);
    }{1/below}{14/below}
}
\providecommand{\FigureEndpointsInCaseIIIdIIIdIIii}{
    \FigureEndpointsInCaseIIIdIIIdII{
        \draw[thick, black!70, dashed] (p15)--(p2)--(p1);
        \draw[thick, violet] (p1)--(p15);
        \draw[thick, blue] (p14)--(p1) (p2) -- (p3)-- (p4);
    }{2/below}{1/below}
}
\providecommand{\FigureEndpointsInCaseIIIdIIIdIIiii}{
    \FigureEndpointsInCaseIIIdIIIdII{
        \draw[thick, red, dashed] (p15)--(p3);
        \draw[thick, violet] (p2)--(p15);
        \draw[thick, blue] (p14)--(p1)--(p2) (p3)-- (p4);
    }{3/below}{2/below}
}

\providecommand{\FigureEndpointsInCaseIIIdIVdI}{
\begin{tikzpicture}[scale=0.8]
\coordinate (p1) at (0.00,0.00);
\coordinate (p2) at (-0.82,0.12);
\coordinate (p3) at (-0.32,2.49);
\coordinate (p4) at (1.83,3.03);
\coordinate (p5) at (0.88,2.39);
\coordinate (p6) at (2.29,2.49);
\coordinate (p7) at (2.43,1.40);
\coordinate (p8) at (1.53,1.93);
\coordinate (p9) at (1.65,1.08);
\coordinate (p10) at (2.23,0.31);
\coordinate (p11) at (1.42,0.39);
\coordinate (p12) at (0.77,0.00);
\coordinate (p13) at (1.69,-0.42);
\coordinate (p14) at (-0.76,-0.52);
\coordinate (p15) at (-1.88,0.26);
\coordinate (p16) at (-1.15,0.85);
\coordinate (p17) at (-2.16,1.77);
\coordinate (p18) at (-1.29,2.68);
\coordinate (p19) at (-0.42,3.07);
\coordinate (p20) at (-0.98,2.02);

\filldraw[gray!10] (p1)--(p2)--(p16)--(p20)--(p3)--(p5)--(p8)--(p9)--(p11)--(p12)--cycle;
\filldraw[violet!20, opacity=0.4] (p1) -- (p2) -- (p3) -- (p4) [bend left=90]to (p13) -- (p12) -- cycle;

\draw[thick, blue] (p1) -- (p2) -- (p3) -- (p4) -- (p5) -- (p6) -- (p7) -- (p8) -- (p9) -- (p10) -- (p11) -- (p12) -- (p13);
\draw[blue] (p13) -- (p14) -- (p15) -- (p16) -- (p17) -- (p18) -- (p19) -- (p20);

\draw[very thick, violet, dashed] (p1) -- (p2) -- (p3) -- (p4) [bend left=90]to (p13) -- (p12) -- cycle;
\foreach \i in {1,...,20}
    \filldraw[black!70] (p\i) circle (1pt);

\foreach \i/\dir/\name in {
    1/above/$s$,
    4/below/$x$,
    10/above/$v$,
    12/above/$s'$,
    13/above/$u$,
    20/below/$t$
}{
    \fill[black] (p\i) circle (1.5pt);
    \node[\dir] at (p\i) {\small \name};
}
\node[violet] at ($(p10)+(0.4,-0.3)$) {$\Gamma$};
    
\end{tikzpicture}
}

\providecommand{\FigureEndpointsInCaseIIIdIVdIdI}[1]{
\begin{tikzpicture}[scale = 0.7]
\coordinate (p1) at (0.00,0.00);
\coordinate (p2) at (-0.82,0.12);
\coordinate (p3) at (-0.32,2.49);
\coordinate (p4) at (1.83,3.03);
\coordinate (p5) at (0.88,2.39);
\coordinate (p6) at (2.29,2.49);
\coordinate (p7) at (2.43,1.40);
\coordinate (p8) at (1.53,1.93);
\coordinate (p9) at (1.65,1.08);
\coordinate (p10) at (2.23,0.31);
\coordinate (p11) at (1.42,0.39);
\coordinate (p12) at (0.77,0.00);
\coordinate (p13) at (1.69,-0.42);
\coordinate (p14) at (-0.76,-0.52);
\coordinate (p15) at (-1.88,0.26);
\coordinate (p16) at (-1.15,0.85);
\coordinate (p17) at (-2.16,1.77);
\coordinate (p18) at (-1.29,2.68);
\coordinate (p19) at (-0.42,3.07);
\coordinate (p20) at (-0.98,2.02);

\filldraw[gray!10] (p1)--(p2)--(p16)--(p20)--(p3)--(p5)--(p8)--(p9)--(p11)--(p12)--cycle;

\draw[thick, blue] (p1) -- (p2) -- (p3) -- (p4) -- (p5) -- (p6) -- (p7) -- (p8) -- (p9) -- (p10) (p11) -- (p12);
\draw[blue] (p13) -- (p14) -- (p15) -- (p16) -- (p17) -- (p18) -- (p19) -- (p20);
#1

\foreach \i in {1,...,20}
    \filldraw[black!70] (p\i) circle (1pt);

\foreach \i/\dir/\name in {
    1/below/$s$,
    10/above/$v$,
    11/below/$v'$,
    12/below/$s'$,
    13/right/$u$
}{
    \fill[black] (p\i) circle (1.5pt);
    \node[\dir] at (p\i) {\small \name};
}  
\end{tikzpicture}
}
\providecommand{\FigureEndpointsInCaseIIIdIVdIdIi}{
    \FigureEndpointsInCaseIIIdIVdIdI{
        \draw[thick, black!70, dashed] (p1)--(p11); 
        \draw[thick, blue] (p10)--(p11);
        \draw[thick, violet](p12)--(p13);
    }
}
\providecommand{\FigureEndpointsInCaseIIIdIVdIdIii}{
    \FigureEndpointsInCaseIIIdIVdIdI{
        \draw[thick, black!70, dashed] (p10)--(p13); 
        \draw[thick, blue] (p1)--(p11);
        \draw[thick, violet](p12)--(p13);
    }
}
\providecommand{\FigureEndpointsInCaseIIIdIVdIdIiii}{
    \FigureEndpointsInCaseIIIdIVdIdI{
        \draw[thick, blue] (p1)--(p11) (p10)--(p13);
    }
}

\providecommand{\FigureEndpointsInCaseIIIdIVdIdII}[1]{
\begin{tikzpicture}[scale=0.5]
\coordinate (p1) at (0.00,0.00);
\coordinate (p2) at (-0.81,0.54);
\coordinate (p3) at (0.63,3.14);
\coordinate (p4) at (1.04,2.56);
\coordinate (p5) at (2.22,2.98);
\coordinate (p6) at (3.19,1.85);
\coordinate (p7) at (2.36,1.72);
\coordinate (p8) at (3.04,0.56);
\coordinate (p9) at (2.18,-0.43);
\coordinate (p10) at (2.25,1.02);
\coordinate (p11) at (1.91,2.29);
\coordinate (p12) at (1.82,0.35);
\coordinate (p13) at (1.02,-0.20);
\coordinate (p14) at (1.15,-0.72);
\coordinate (p15) at (-0.59,-0.55);
\coordinate (p16) at (-1.96,0.95);
\coordinate (p17) at (-0.98,2.69);
\coordinate (p18) at (-1.01,1.36);
\coordinate (p19) at (-0.54,2.00);

\filldraw[gray!10] (p1)--(p2)--(p18)--(p19)--(p4)--(p11)--(p10)--(p12)--(p13)--cycle;
\draw[thick, blue] (p1) -- (p2) -- (p3) -- (p4) -- (p5) -- (p6) -- (p7) -- (p8) -- (p9) -- (p10) -- (p11) (p12) -- (p13) -- (p14);
\draw[blue] (p14) -- (p15) -- (p16) -- (p17) -- (p18) -- (p19);
#1

\foreach \i in {1,...,19}
    \filldraw[black!70] (p\i) circle (1pt);

\foreach \i/\dir/\name in {
    1/below/$s$,
    13/right/$s'$,
    10/right/$v'$,
    11/right/$p$,
    12/above left/$q$,
    4/below/$p'$
}{
    \fill[black] (p\i) circle (1.5pt);
    \node[\dir] at (p\i) {\small \name};
}  
\end{tikzpicture}
}
\providecommand{\FigureEndpointsInCaseIIIdIVdIdIIi}{
    \FigureEndpointsInCaseIIIdIVdIdII{
        \draw[thick, black!70, dashed] (p1)--(p12);
        \draw[thick, blue] (p11)--(p12);
    }
}
\providecommand{\FigureEndpointsInCaseIIIdIVdIdIIii}{
    \FigureEndpointsInCaseIIIdIVdIdII{
        \draw[thick, blue] (p1)--(p12);
    }
}

\providecommand{\FigureEndpointsInCaseIIIdIVdII}[1]{
\begin{tikzpicture}[scale=0.5]
\coordinate (p1) at (0.00,0.00);
\coordinate (p2) at (-0.83,0.37);
\coordinate (p3) at (-1.24,1.17);
\coordinate (p4) at (0.60,2.46);
\coordinate (p5) at (2.60,1.18);
\coordinate (p6) at (2.05,0.23);
\coordinate (p7) at (1.29,0.03);
\coordinate (p8) at (1.29,-0.76);
\coordinate (p9) at (-0.21,-0.71);
\coordinate (p10) at (-1.75,0.44);
\coordinate (p11) at (-1.55,2.48);
\coordinate (p12) at (-0.86,1.75);
\coordinate (p13) at (-0.37,2.15);
\coordinate (p14) at (-0.27,2.96);
\coordinate (p15) at (1.43,2.93);
\coordinate (p16) at (1.97,1.96);
\coordinate (p17) at (2.44,2.61);
\coordinate (p18) at (3.22,1.22);
\coordinate (p19) at (2.65,-0.18);
\coordinate (p20) at (2.49,0.62);

\draw[thick, blue] (p1) -- (p2) -- (p3) (p4) -- (p5) -- (p6) -- (p7) -- (p8) -- (p9) -- (p10) -- (p11) -- (p12) -- (p13) -- (p14) -- (p15) -- (p16) -- (p17) -- (p18) -- (p19) -- (p20);
#1
\foreach \i in {1,...,20}
    \filldraw[black!70] (p\i) circle (1pt);

\foreach \i/\dir/\name in {
    1/below/$s$,
    7/above/$s'$,
    3/right/$u$,
    12/above/$u'$,
    4/right/$v$
}{
    \fill[black] (p\i) circle (1.5pt);
    \node[\dir] at (p\i) {\small \name};
}  
\end{tikzpicture}
}
\providecommand{\FigureEndpointsInCaseIIIdIVdIIi}{
    \FigureEndpointsInCaseIIIdIVdII{
        \draw[thick, black!70, dashed] (p1)--(p4);
        \draw[thick, blue] (p3)--(p4);
    }
}
\providecommand{\FigureEndpointsInCaseIIIdIVdIIii}{
    \FigureEndpointsInCaseIIIdIVdII{
        \draw[thick, blue] (p1)--(p4);
    }
}

\begin{document}

\maketitle

\begin{abstract}
    Consider a crossing-free polygonal path with $n$ vertices and a flip operation that removes an edge and adds another one,  without changing the vertex set and keeping the path crossing free. It is conjectured that any two such paths can be transformed into each other by a sequence of flips. Recently, Kleist et al. (2024) proved the conjecture for the special cases of vertices arranged in two convex layers. However, the number of flips in their proof may be exponential. In this paper, we show that $O(n^2)$ flips suffice. More precisely, for a two-layer point set $S=L_0\cup L_1$, where $L_0$ and $L_1$ denote the outer and inner layers, respectively, our method uses $O(|L_0|+|L_1|^2)$ flips.
\end{abstract}

\section{Introduction}

\begin{figure}[ht]
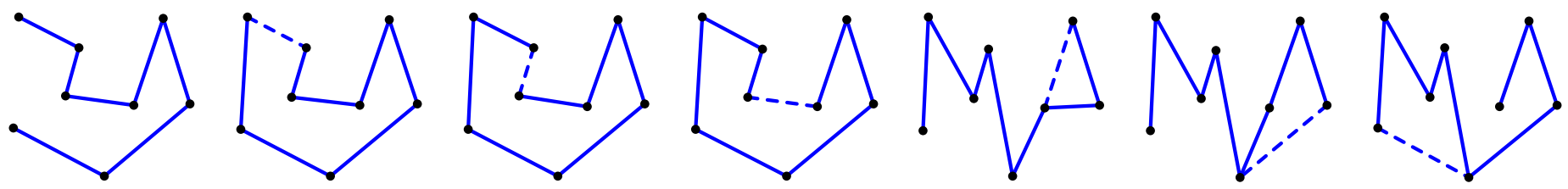

        \centering
            \Example
        \caption{A flip sequence connecting a pair of paths}
        \label{fig:example}
\end{figure}

Reconfiguration studies a family of configurations that are changed into each other using a flip operation. In our case, the \emph{configurations} are crossing-free spanning paths of a set $S$ of $n$ points and the \emph{flip} operation removes an edge and adds another one while preserving a valid configuration. 
Given a set $S$, there is an associated \emph{flip graph} whose vertices are configurations and edges are defined by flips. The \emph{diameter} of the flip graph is the number of flips needed, in the worst case, to transform one path into another. We refer to paths in the flip graph as \emph{flip sequences} to avoid confusion. An example of flip sequence is drawn Fig.~\ref{fig:example}.

The problem was introduced by~\cite{RiveraCampoUrrutiaGalicia01} in 2001, together with an implicit \emph{conjecture} that the flip graph is connected and a proof of it for the special cases of points $S$ in \emph{convex} position ($S$ is the set of convex hull vertices). The first polynomial bound on the diameter for the convex case appears in~\cite{AklIslamMeijer07}, at most $2n-5$. Later a tight bound of $2n-5$ for $n \le 4$ and $2n-6$ for $n \ge 5$ has been shown~\cite{ChangWu09}. In the convex case,  the shortest flip sequence can be computed in linear time~\cite{AichholzerDorfer26}.

Proving the conjecture for arbitrary point sets remains challenging and many authors considered several other special cases; we cite the most relevant for this paper. For convex position plus one \emph{interior} point, the diameter is at most $2n-4$~\cite{AichholzerKMOPV23}.  For convex position plus one \emph{exterior} point, the diameter is at most $2n$~\cite{BoucardFR24}. Both cases are special cases of a more general \emph{two-layer} point set $S=L_0 \cup L_1$ where the \emph{outer layer} $L_0$ is the set of vertices in the convex hull of $S$ and the \emph{inner layer} $L_1$ is also in convex position. In~\cite{KleistKR24}, they show that the flip graph for two layers is connected, but the diameter implied by their proof is exponential. In this paper, we prove the following quadratic bound.

\begin{theorem}\label{th:main}
For a two-layer point set $S=L_0\cup L_1$, where $L_0$ and $L_1$ denote the outer and inner layers, the diameter of the flip graph is $O(|L_0|+|L_1|^2)$ and computing a flip sequence connecting a pair of paths takes $O(|L_0|+|L_1|^2)$ time.
\end{theorem}

\section{Canonical Paths and Potential Reduction}\label{sec:Canon}

This section assembles the main elements allowing us to prove Theorem \ref{th:main}. Let $S = L_0\cup L_1$ be a two-layer point set with $L_0$ and $L_1$ the outer and inner layer respectively. Let $n_0 = |L_0|$ and $n_1 = |L_1|$. 

We first prove the connectivity of a particular family of paths that we will call \emph{canonical paths} (Subsection \ref{sub:Recon}).
The second step is to show that any arbitrary plane spanning path on a two-layer point set can be transformed into a canonical path. 
We achieve this goal by defining  a potential function $\Phi$ on paths (subsection \ref{sub:Potential}). The potential has three properties (subsection \ref{sub:Properties}):
\begin{itemize}
    \item the potential of a path is $1$ if and only if that path is canonical
    \item the potential is strictly positive, so that the set of the canonical paths is exactly the class of paths with minimum potential
    \item if the potential of a path is not minimal i.e if the path is not canonical, we provide an algorithm for getting a sequence of flips decreasing the potential by at least $1$. The number of flips required for a potential decreasing is precisely the key explaining the bound on the diameter of the flip graph given in Theorem \ref{th:main}.
\end{itemize}
In other words, a sequence of flips decreasing the potential of a path until the minimum value $1$ transforms the initial path into a canonical path. As all the canonical paths are connected, the connection of any path to a canonical path  proves the connectivity of the whole flip graph. It yields Theorem \ref{th:main}. 

\subsection{Terminology}

Before starting to define the canonical paths and the potential, we introduce some notations.
For two vertices $u,v\in S$ in a path $P$, we say that $u$ \emph{sees} $v$ if the interior of the segment $uv$ does not intersect any other edge of $P$.

To distinguish the neighbors of the vertices in a path, we may orient the path. When a path is oriented, its first and last vertices are called the \emph{start} and \emph{end}, respectively. An \emph{$st$-path} is an oriented path with start $s$ and end $t$. For a vertex $u$, we denote by $\inc(u)$ and $\out(u)$ its incoming and outgoing edges, when they exist. We write $\pred(u)$ and $\succ(u)$ for the predecessor and successor of $u$, respectively. Whenever an edge is written as $uv$ in an oriented path, we implicitly assume that $u=\pred(v)$.

For an $st$-path $P$, the subpath from $u\in S$ to $v\in S$ is denoted by $P[u:v]$. A prefix of the path is any subpath of the form $P[s:u]$, $u\in S$, and a suffix of the path is any path of the form $P[u:t]$, $u\in S$. 

\subsection{Reconfiguration of Canonical Paths}\label{sub:Recon}
\begin{figure}[ht]
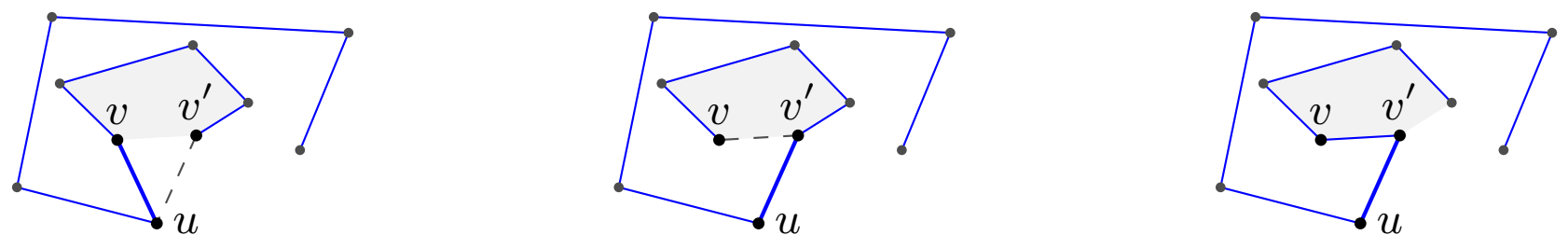

    \centering

    \begin{subfigure}[t]{0.3\textwidth}
        \centering
        \FigureLeastPotentiali
        \label{fig:shift-bridge-1}
    \end{subfigure}
    \hfill
    \begin{subfigure}[t]{0.3\textwidth}
        \centering
        \FigureLeastPotentialii
        \label{fig:shift-bridge-2}
    \end{subfigure}
    \hfill
    \begin{subfigure}[t]{0.3\textwidth}
        \centering
        \FigureLeastPotentialiii
        \label{fig:shift-bridge-3}
    \end{subfigure}

    \caption{Counterclockwise bridge shift on the inner layer}
    \label{fig:shift-bridge}
\end{figure}

A path is \emph{canonical} if it first visits all vertices of one layer in
cyclic order and then all vertices of the other layer in cyclic order. The directions chosen on the two layers do not need to be the same. For instance, all paths in Figure~\ref{fig:shift-bridge} are canonical.

The unique edge of a canonical path connecting an outer point to an inner point is called a \emph{bridge}. A canonical path is completely determined by its bridge, and, for each layer, the direction (clockwise or counterclockwise) in which the path traverses that layer from the corresponding extremity of the bridge.

To reconfigure canonical paths, two useful observation can be made:
\begin{itemize}
    \item The traversal direction of either layer can be reversed in 1 flip.
    \item Either extremity of the bridge can be shifted to their layer neighbor they are not connected with in 1 flip, if it does not introduce crossing.
\end{itemize}

Using these observations, we can reconfigure canonical paths in linear flips while preserving their structure.

\begin{lemma}\label{lem:connected_canonical}
    The subgraph of the flip graph of canonical paths on a two-layer point set is connected with diameter $O(n_0+n_1)$ flips.
\end{lemma}

\begin{proof}
    Let $P$ and $P'$ be two canonical paths with bridges $uv$ and $u'v'$, where $u,u'\in L_0$ and $v,v'\in L_1$.

    Consider a triangulation of the ring between $L_0$ and $L_1$ that contains both $uv$ and $u'v'$ and such that and no triangle connects three points of $L_0$. In such a triangulation, the edges joining $L_0$ and $L_1$ form a cyclic sequence $T$ where each consecutive edge differ only by one endpoint.

    Fix a direction along this sequence, say counterclockwise. First, if necessary, flip the edges adjacent to the bridge to connect $u$ and $v$ to $\ccw(u)$ and $\ccw(v)$ respectively.

    Starting from $uv$, repeatedly flip the bridge to the next one of $T$ in counterclockwise direction, and do another flip to connect the extremity of the bridge that shifted to its clockwise neighbor again. An Illustration of these two flips is given in Figure~\ref{fig:shift-bridge}
    This second flip preserves the loop invariant that the path is always a canonical path in which both extremities of the bridge are connected to their clockwise neighbor.

    Repeating this process eventually transforms the bridge $uv$ into $u'v'$. Finally, at most two additional flips are needed to adjust the traversal directions on the two layers so that the resulting canonical path is exactly $P'$
    
    At the end of that loop we can correct the orientation of the path at the extremity of the bridge to match $P'$ in 2 flips. Therefore $P$ can be transformed into $P'$. 
    
    Moreover, the sequence $T$ contains exactly $n_0+n_1$ edges. So the whole process uses at most $2(n_0+n_1)+4$ flips, proving the claim. \qed
\end{proof}

\subsection{Potential}\label{sub:Potential}

We now introduce the potential used for connecting any path to a canonical path.
Its value is based on a classification of edges according to their position with respect to the two layers. We distinguish five types of edges:
\begin{itemize} 
    \item \emph{level edges}, joining two consecutive vertices on the same layer; 
    \item \emph{bridges}, joining an outer and an inner vertex without crossing the convex hull of $L_1$; 
    \item \emph{small diagonals}, joining two non-consecutive vertices of $L_1$; 
    \item \emph{medium diagonals}, joining an outer and an inner vertex through the interior of the convex hull of $L_1$; 
    \item \emph{big diagonals}, joining two non-consecutive vertices of $L_0$. 
\end{itemize}

\begin{figure}[ht]
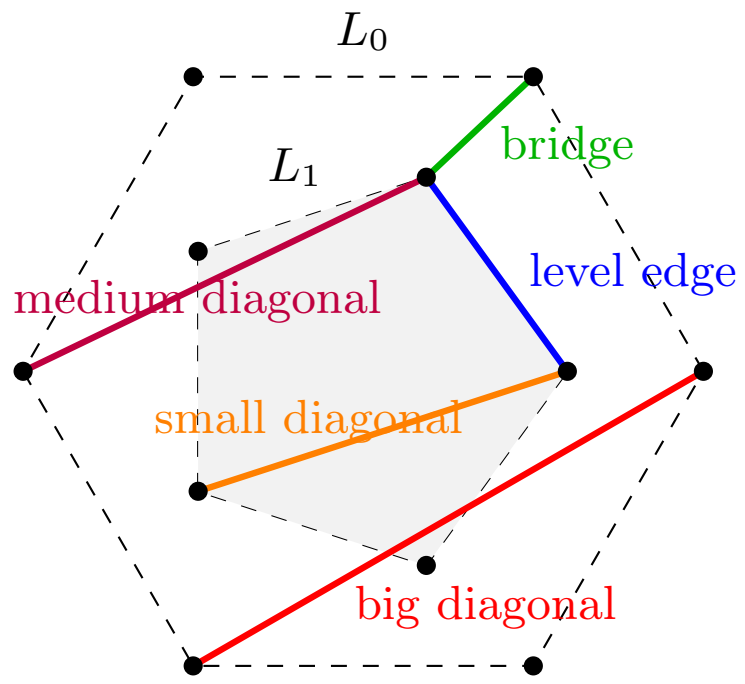

\centering
\FigureEdgeTypes
\caption{Illustration of the five edge types in a two-layer point set.}
\label{fig:edge-types}
\end{figure}

These edge types are represented in Figure~\ref{fig:edge-types}.

We assign a weight to each edge type and define the potential $\Phi(P)$ of a path $P$ as
\begin{equation}\label{eq:potential}
    \Phi(P) =  0L(P) + 1B(P) + 1S(P) + 3M(P) + 4G(P)
\end{equation}
where $L(P)$, $B(P)$, $S(P)$, $M(P)$ and $G(P)$ denote the numbers of level edges, bridges, small diagonals, medium diagonals and big diagonals in $P$, respectively. The coefficients in~\eqref{eq:potential} are chosen so that canonical paths are exactly the paths of minimum potential.

\subsection{Potential Properties and Proof of Theorem \ref{th:main}}\label{sub:Properties}

The first properties are straightforward. First, by definition, the potential is a strictly positive integer. Secondly, paths of minimal potential are canonical:

\begin{lemma}\label{lem:least-potential-path} 
    The minimum value of $\Phi$ is $1$, and a path has potential $1$ if and only if it is canonical. \end{lemma}
    
\begin{proof}
    Every spanning path must contain an edge joining the two layers. Such an edge is either a bridge or a medium diagonal, both of which contribute at least $1$ to the potential. Therefore $\Phi(P)\ge 1$ for every path $P$.

    Now let $P$ be a path with $\Phi(P)=1$. Since every edge type except level edges has weight at least $1$, the path must contain exactly one bridge and all its remaining $n_0+n_1-2$ edges must be level edges.

    Consequently, the restriction of $P$ to $L_0$ contains exactly $n_0-1$ level edges and the restriction of $P$ to $L_1$ contains exactly $n_1-1$ level edges. Hence the vertices of each layer are traversed consecutively in cyclic order. The unique bridge connects the two layer-traversals, so $P$ is a canonical path.
    
    Conversely, every canonical path consists of a single bridge and $n_0+n_1-2$ level edges, and therefore has potential $1$. \qed
\end{proof}

\begin{lemma}\label{P2}
If a path is not canonical, we can reduce its potential by at least $1$ with a sequence of at most $O(n_1)$ flips.
\end{lemma}

\begin{lemma}\label{lem:not-too-many-hard-cases}
Throughout the process of diminishing the potential of the path, at most $O(n_1)$ potential reduction steps require more than 1 flip.
\end{lemma}

These last two lemmas enable the proof of Theorem~\ref{th:main}.

\begin{proof}[Theorem~\ref{th:main}]
    Let $P$ be a path. By Lemma~\ref{P2}, we can diminish its potential in at most $O(n_1)$ flips, until it has least potential and becomes canonical (Lemma~\ref{lem:least-potential-path}). Among each of these reduction steps, at most $O(n_1)$ of them may require more than 1 flip to be performed, according to Lemma~\ref{lem:not-too-many-hard-cases}. Therefore, since $\Phi(P) = O(n_0+n_1)$, this means that the reconfiguration of $P$ into a canonical path requires at most $O(n_0+n_1^2)$.

    Moreover, using Lemma~\ref{lem:connected_canonical}, we can reconfigure any canonical paths into one another in $O(n_0+n_1)$ flip, meaning we can reconfigure any path into one another in $O(n_0+n_1^2)$ flips.\qed
\end{proof}

\begin{figure}[ht]
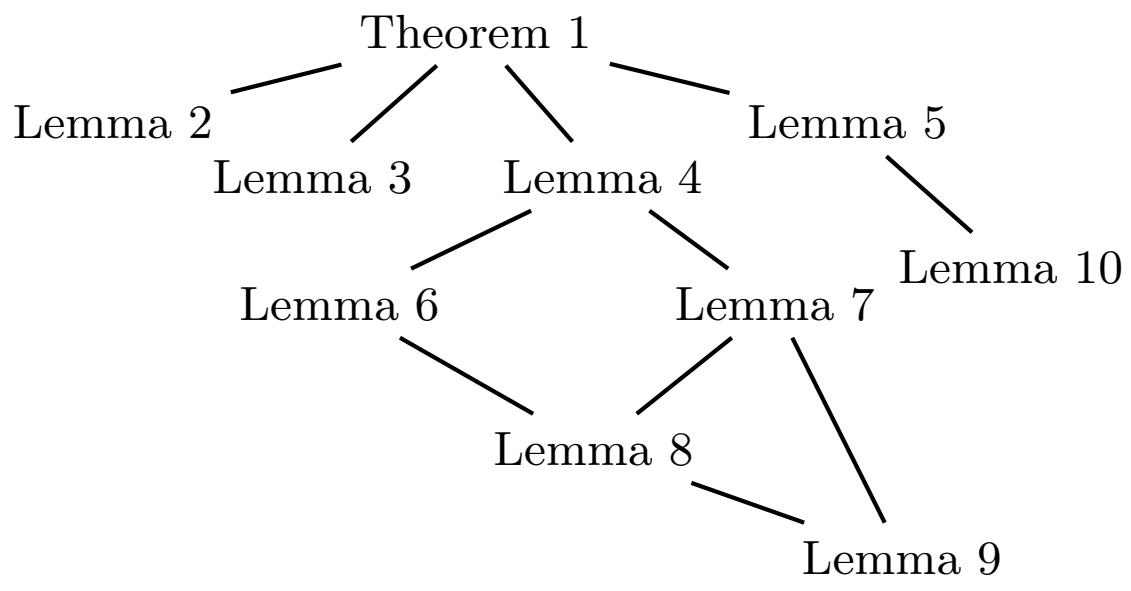

        \centering
            \Tree
        \caption{The tree of dependencies of the different lemmas allowing us to prove Theorem \ref{th:main}}
        \label{fig:tree}
\end{figure}

The later material in Sections ~\ref{sec:PotentialReduction}, \ref{lem:endpoint-in} and \ref{sec:not-too-many-hard-cases} are respectively devoted to the proofs of Lemma~\ref{P2} and Lemma~\ref{lem:not-too-many-hard-cases}. An organization of the dependencies between the previous and next claims is illustrated in Fig.~\ref{fig:tree}

\section{Potential Reduction}\label{sec:PotentialReduction}

In this section, we provide a sketch of proof of Lemma \ref{P2}. 
The different procedures constituting the algorithm leading to this lemma and their respective proof of correction are given in Section~\ref{sec:algo-reduction}.
 
To this end, we describe an algorithm that, given a path $P$ with $\Phi(P)>1$, performs a sequence of flips that decreases the potential.
The algorithm uses two different procedures, depending on whether an endpoint of $P$ is on the outer layer or if both are in the inner layer.
We give a high level description of these procedures.

The procedure \textsc{OuterReduction} takes as input an oriented path whose start vertex lies on the outer layer. Its goal is to create a new level edge incident to the start. When such an edge cannot be created immediately, it tries the same thing with the end. If this fails again, the procedure uses its end vertex to change the orientation of the path next to the start vertex, so that a new level edge incident to the start can be created. This procedure performs a constant flip sequence.

\begin{lemma}\label{lem:endpoint-out}
Let $P$ be an $st$-path where the start $s\in L_0$ is an outerpoint. Then the procedure \textsc{OuterReduction} reduces the potential of the path in $O(1)$ flips.
\end{lemma}

The procedure \textsc{InnerReduction} handles the case where both endpoints lie on the inner layer. As before, the primary objective is to create a new level edge incident to the start vertex. If this fails, depending on the type of the incoming edge of the layer neighbor of the start, the procedure either performs a direct reduction or transforms the path into a simpler configuration that admits further reductions. This procedure may perform $O(n_1)$ flips in certain cases.

\begin{lemma}\label{lem:endpoint-in}
Let $P$ be an $st$-path where both endpoints are in $L_1$. Then the procedure \textsc{InnerReduction} reduces the potential of the path in $O(n_1)$ flips.
\end{lemma}

The two lemmas \ref{lem:endpoint-out} and \ref{lem:endpoint-in} directly imply Lemma~\ref{P2}.
Their two procedures \textsc{OuterReduction} and \textsc{InnerReduction} may invoke some common subprocedures \textsc{ObstrReduction} and \textsc{DiagObstrReduction} when common cases arise. These two subprocedures handle cases where
some visibility between some points are obstructed by some edges of the path. These cases always involve edges with heavy weight, so the goal of these subprocedures is to find a way to flip them. 

\begin{lemma}\label{lem:correct_obstr}
Let $P$ be a path starting at $s\in L_1$ and let $q$ be a neighbor not seen by $s$ because some edge of $P$ blocks the visibility between them.
Then the procedure \textsc{ObstrReduction} reduces the potential of the path in $O(1)$ flips.
\end{lemma}

\begin{lemma}\label{lem:correct_diagobstr}
Let $P$ be a path starting at $s\in L_1$. Suppose that $P$ contains a diagonal $uv$, where either $uv$ is a big diagonal or $uv$ is a medium diagonal with $u$ a cyclic neighbor of $s$ in $L_1$. Assume that $s$ sees $u$ but not $v$, and that there exists an inner vertex $p\in\triangle suv$ that sees both $s$ and $v$. Moreover, assume that the concave quadrilateral $suvp$ is empty. 

Let $u'$ be the last outer vertex on $P[s:u]$ and let $v'$ be the first neighbor of $v$ visited by $P$. Assume further that 
\begin{enumerate} 
\item $P[s:p]$ lies entirely in $L_1$ and visits its vertices in cyclic order from $s$ to $p$; 
\item and $P[p:u']$ visits the remaining inner vertices between $s$ and $p$, together with the outer vertices from $v'$ to $u'$, preserving their cyclic order on $L_0$. 
\end{enumerate} 

Then \textsc{DiagObstrReduction} reduces the potential of $P$ using $O(1)$ flips.
\end{lemma}

We detail the procedures and prove Lemmas~\ref{lem:endpoint-out} to~\ref{lem:correct_diagobstr} in the following section.

\section{Procedures reducing the potential of a path}\label{sec:algo-reduction}
In this section, we provide the different procedures reducing the potential of a path, and prove their correctness and their complexity. We start by presenting the two main procedures, \textsc{OuterReduction} (Section~\ref{sec:outer-reduction}) and \textsc{InnerReduction} (Section~\ref{sec:inner-reduction}), and finish by the last two sub procedures \textsc{ObstrReduction} and \textsc{DiagObstrReduction} (Section~\ref{sec:obstr-reduction}).

\subsection{Procedure \textsc{OuterReduction} and proof of Lemma~\ref{lem:endpoint-out}}\label{sec:outer-reduction}

\textbf{Procedure} \textsc{OuterReduction}.

\textit{Input:}
a path $P$ whose start $s\in L_0$ lies on the outer layer, 

and a neighbor $s'$ of $s$ that does not share an edge with it. 

Assume $s'=\ccw(s)$ by symmetry.

Let $t$ be the end of the path.

\begin{enumerate}
\item If $s'=t$, pick any non-level edge and flip it to $st$.

\item Else if the incoming edge of $s'$ is not a level edge, flip it to $ss'$.

\item Otherwise: Let $t'$ be a neighbor of $t$ that does not share an edge with it.
    \begin{enumerate}
    \item If $t'$ does not see $t$, reverse the order of the path and apply the \textsc{ObstrReduction} procedure on that obstruction.
    
    \item Else if the outgoing edge of $t'$ is not a level edge, flip it to $tt'$.

    \item Else if $t'\in P[s:s']$, perform
    \[\operatorname{out}(t')\longrightarrow t't, \qquad \operatorname{in}(s')\longrightarrow ss'\]

    \item Otherwise, 
    
    Let $uv$, with $u = \pred(v)$, be the last small diagonal in $P[\succ(s'):t]$, and let $v'$ be neighbor of $v$ not seen by $t'$.
    Then perform $uv\longrightarrow ut$, and apply recursively \textsc{OuterReduction} using $v'$ for the non connected neighbor of the new start $v$.
    \end{enumerate}

\end{enumerate}

\begin{proof}[Lemma~\ref{lem:endpoint-out}]
    Let $P$ be an $st$-path starting at $s\in L_0$, and assume $|L_0|\geq 3$, otherwise the lemma is trivial.  
Let $s'\in L_0$ be one of the cyclic neighbor of $s$ and assume $s' = \ccw(s)$ by symmetry. Since these points are neighbors of the outer layer, they necessarily see each other.

Therefore the flips in Cases~1 and~2 are legal and decrease the potential. 

So now suppose the premise of Case~3 is verified: the incoming edge of $s'\neq t$ is a level edge. Then the path has a particular structure.
Indeed, as illustrated in Figure~\ref{fig:outer-path-structure}, the prefix $P[s:s']$ wraps around the suffix $P[\succ(s'):t]$ so that the latter is forced to stay confined in $L_1$. In particular, the end $t\in L_1$ is an inner point. This restriction on the suffix means that $P[s:s']$ must visit all outer points, and we can prove that all the points from $L_0$ must appear in a cyclic order in that prefix.
\begin{figure}[ht]
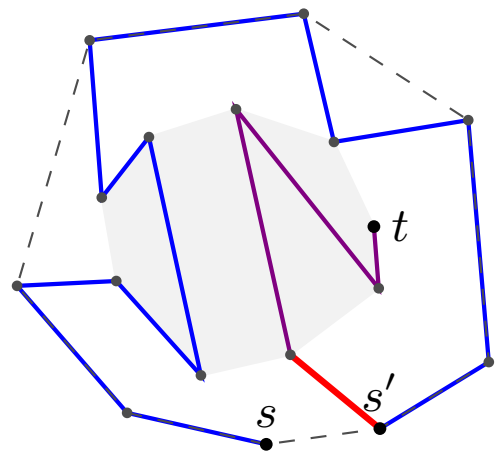

    \centering
    \FigureEndpointOutPrefixStructure
    \caption{Structure of the path in Case~3 of procedure \textsc{OuterReduction}}
    \label{fig:outer-path-structure}
\end{figure}

Also, since the incoming edge of $s'$ is a level edge, its outgoing edge is not as $s'$ does not share an edge with $s$. So the remaining cases try to reverse the orientation of the path in $s'$ if it cannot be immediately be reduced near $t$.

Let $t'\in L_1$ be a neighbor of $t$ such that $t'$ does not share an edge with $t$. 
Such a $t'$ always exists unless $|L_1|\le 2$ and the path already has potential~1.
We distinguish cases depending on visibility between $t$ and $t'$ and the type of the outgoing edge of $t'$.

\begin{itemize}
    \item \textbf{Case~3.(a)}: If $t$ does not see $t'$, then applying \textsc{ObstrReduction} reduces the potential in $O(1)$ according to Lemma~\ref{lem:correct_obstr}.
    \item \textbf{Case~3.(b)}: If $t$ sees $t'$ and the outgoing edge of $t'$ is not a level edge, then $\out(t')\to t't$ creates a new level edge, diminishing the potential.
\end{itemize}

Now let's suppose that $t$ sees $t'$, and $\out(t')$ is a level edge. 
\begin{itemize}
    \item \textbf{Case 3.(c)}: If $t'\in P[s:s']$ or equivalently $s'\in P[t':t]$, the first flip ($\out(t') \to t't$) is legal and does not change the potential. The second flip $\inc(s')\to ss'$ is also legal as $s'$ sees $s$. Moreover, after the first flip the orientation of the path in $s'\in P[t':t]$ is reversed, exchanging $\inc(s')$ with the non-level edge $\out(s')$. Therefore the second flip flips a non-level edge to a level edge, which diminishes the potential. The flip sequence produced is illustrated in Figure~\ref{fig:outer-3-3-1}
    \begin{figure}[ht]
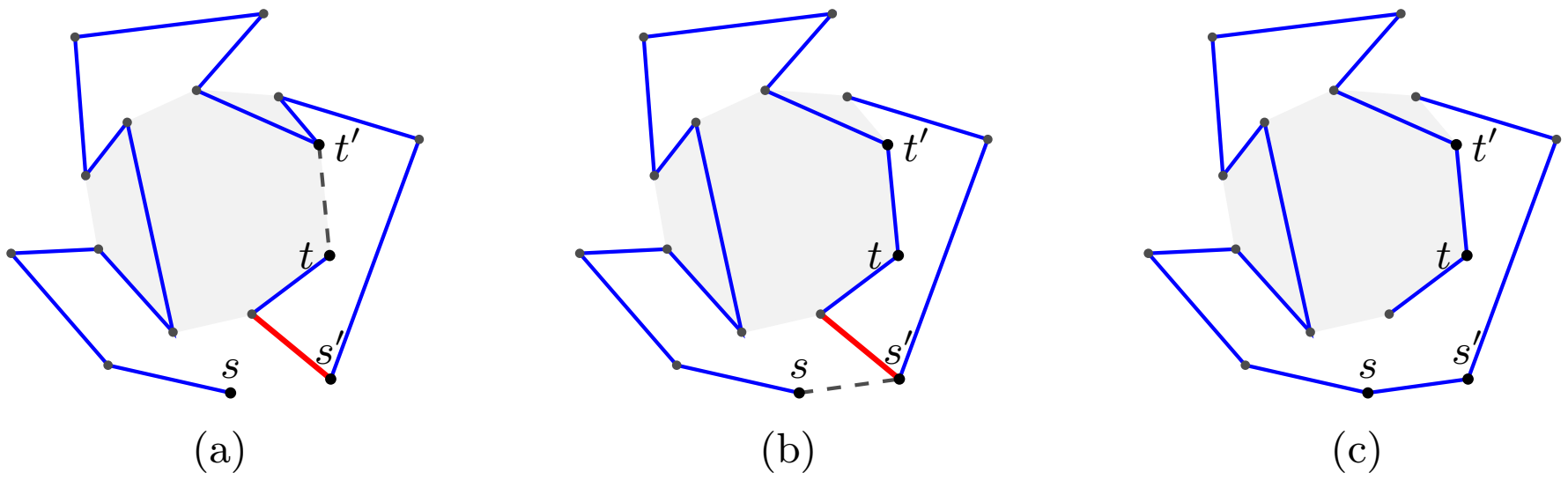

    \centering
    \begin{subfigure}[t]{0.3\textwidth}
        \centering
        \FigureEndpointOutCaseIIIdIIIdIi
        \caption{}
        \label{fig:outer-3-3-1_1}
    \end{subfigure}
    \hfill
    \begin{subfigure}[t]{0.3\textwidth}
        \centering
        \FigureEndpointOutCaseIIIdIIIdIii
        \caption{}
        \label{fig:outer-3-3-1_2}
    \end{subfigure}
    \hfill
    \begin{subfigure}[t]{0.3\textwidth}
        \centering
        \FigureEndpointOutCaseIIIdIIIdIiii
        \caption{}
        \label{fig:outer-3-3-1_3}
    \end{subfigure}

    \caption{Reduction of potential in Case~3.(c) of Procedure \textsc{OuterReduction}.}
    \label{fig:outer-3-3-1}
\end{figure}

    \item \textbf{Case 3.(d)}: Suppose $t'\in P[\succ(s'):t]$. In this case, we can show that $t' = \succ(s')$. Indeed, if that was not the case, then as $P[\succ(s'):t]\subset L_1$ and $\out(t')$ is a level edge, $\inc(t')$ must be a small diagonal. This diagonal would cut $L_1$ in two parts, separating $\succ(t')$ and $t$ the two neighbors of $t'$ and introducing a crossing with the suffix $P[\succ(t'):t]$. Therefore in this case $t' = \succ(s')$.

    From this observation we can deduce that $P[t':t]$ necessarily contains a small diagonal, as if it was not the case, $P[t':t]$ would visit all inner points in cyclic order, which would imply that the path is already canonical.

    Now, as in the procedure, let $uv$ be the last diagonal of $P[t':t]$. Then it splits $L_1$ in half and we can define $v'$ to be the neighbor of $v$ not seen by $t'$. The flip $uv\to ut$ does not necessarily decrease the potential, but it changes the end of the path to $v$, so that applying again the procedure with $v'$ for the choice of the neighbor of $v$ will not lead to that case again, since $v'\neq t$ so $v' \neq \succ(s')$. And since all the other cases decrease the potential in $O(1)$ flips, this one does as well.
    The flip sequence produced in this case is illustrated in Figure~\ref{fig:outer-3-3-2}
\end{itemize}
\begin{figure}[ht]
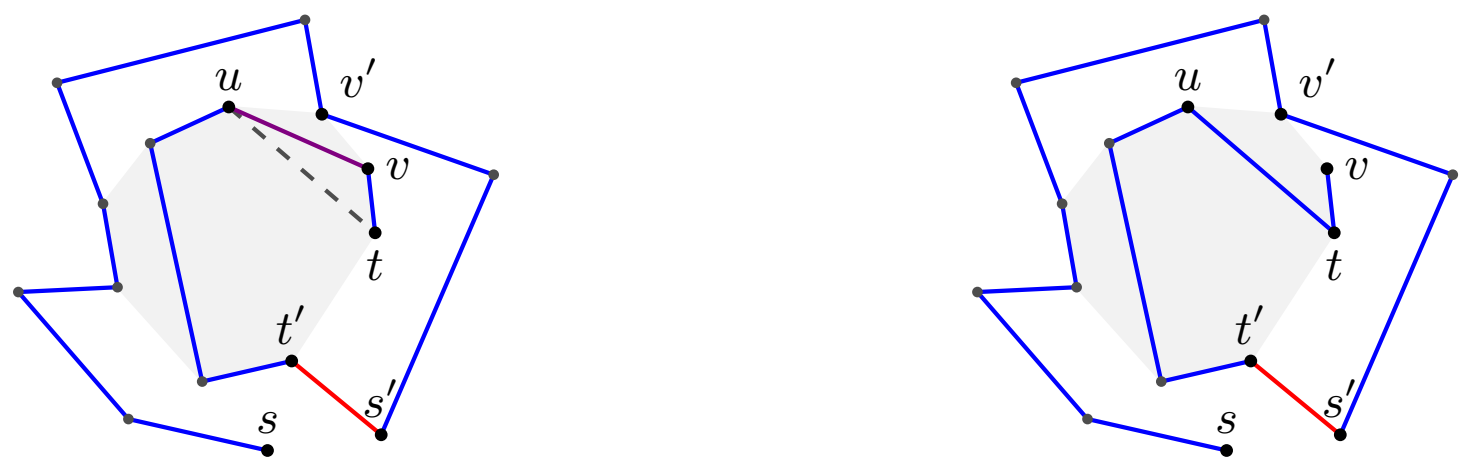

    \centering
    \begin{subfigure}[t]{0.48\textwidth}
        \centering
        \FigureEndpointOutCaseIIIdIIIdIIi
        \label{fig:outer-3-3-2_1}
    \end{subfigure}
    \hfill
    \begin{subfigure}[t]{0.48\textwidth}
        \centering
        \FigureEndpointOutCaseIIIdIIIdIIii
        \label{fig:outer-3-3-2_2}
    \end{subfigure}

    \caption{Reduction Case~3.(d) to other cases of procedure \textsc{OuterReduction}}
    \label{fig:outer-3-3-2}
\end{figure}

This proves by case analysis that the procedure \textsc{OuterReduction} can always reduce the potential of a non-canonical path with an outer endpoint in $O(1)$ flips. \qed
\end{proof}

\subsection{Procedure \textsc{InnerReduction} and proof of Lemma~\ref{lem:endpoint-in}}\label{sec:inner-reduction}

\textbf{Procedure} \textsc{InnerEndpointsReduce}.
\textit{Input:}
a path $P$ whose start $s\in L_1$ and end $t\in L_1$ lie on the inner layer, 

and a neighbor $s'$ of $s$ that does not share an edge with it. 

Assume $s'=\ccw(s)$ by symmetry.
\begin{enumerate}
\item If $s$ does not see $s'$, apply the \textsc{ObstrReduction} procedure on that obstruction. 

\item Else if the incoming edge of $s'$ is not a level edge, flip it to $ss'$.

\item Else if $s' = t$, pick any non-level edge and flip it to $ss'$.

\item Else if the outgoing edge of $s'$ is a small diagonal: 
Let $x = \succ(s')$. Repeat until the potential decreases:
let $s$ be the start of the path, and $s' = \ccw(s)$, and perform
\[s'x\to sx, \qquad \operatorname{in}(s)\to s's.\]

\item Else if the outgoing edge of $s'$ is a medium diagonal:
Let $x = \succ(s')$. 
If $s$ and $t$ are on the same side of $s'x$ in $L_1$, flip $\ccw(s')s'\to ss'$, rename $\ccw(s')$ as $s$, and take the reflect the path so that $s' = \ccw(s)$ again.
Then
\begin{enumerate}
    \item If $x$ sees $s$, repeat until the potential decreases or $s$ no longer sees $x$:
let $s$ be the start of the path, and $s' = \ccw(s)$, and perform
\[s'x\to sx, \qquad \operatorname{in}(s)\to s's.\]

    If the potential has not decreased, then follow the procedure \textsc{DiagObstrReduction} with the start $s$ and diagonal $s'x$.

    \item If $x$ does not see $s$, follow the procedure \textsc{DiagObstrReduction} with the start $s$ and diagonal $s'x$.
\end{enumerate}

\item Otherwise, the outgoing edge of $s'$ is a bridge.
\begin{enumerate}
    \item If $P[s:s']\not\subset L_1$,
    Let $u=\succ(s')$.

    If $s'$, $u$, $\succ(u)$ are not in clockwise order, flip $\ccw(s')s'$ to $ss'$, rename $\ccw(s')$ as $s$, and reflect the path so that $s'$ becomes $\ccw(s)$ again, and the order of $s'$, $u$, $\succ(u)$ becomes clockwise.

    Let $v=\ccw(u)$ and $v'=\succ(v)$.
    \begin{enumerate}
        \item If $s$ sees $v'$, perform $vv'\to sv'$, then $s'u\to vu$.
        \item Else let $pq$ be the last small diagonal of $P[v':s']$, with $p=\pred(q)$, and let $p'=\ccw(p)$.
        Then flip $pq\to sq$ and recursively apply \textsc{InnerEndpointsReduce} using $p'$ as the neighbor of the new start $p$.
    \end{enumerate}
    
    \item Otherwise $P[s:s']\subset L_1$.
    Let $uv$ be the first small diagonal of $P[s:s']$, with $u=\pred(v)$.
    let $u'$ be the neighbor of $u$ not seen by $s'$.
    Then flip $uv$ to $sv$, and recursively apply \textsc{InnerEndpointsReduce} using $u'$ as the neighbor of the new start $u$.
\end{enumerate}

\end{enumerate}

\begin{proof}[Lemma~\ref{lem:endpoint-in}]
Let $P$ be an $st$-path with both endpoints on the inner layer. 
Let $s'\in L_1$ be a cyclic neighbor of $s$ that does not share an edge with $s$, and assume $s'=\ccw(s)$ by symmetry.

In the first three cases we try to create a new level edge between $s$ and $s'$.
\begin{itemize}
    \item \textbf{Case~1:} If $s$ does not see $s'$, then \textsc{ObstrReduction} reduces the potential in $O(1)$ according to Lemma~\ref{lem:correct_obstr}.
    \item \textbf{Case~2:} If $s$ sees $s'$ and the incoming edge of $s'$ is a non level edge, then flipping it to $ss'$ instantly decreases the potential
    \item \textbf{Case~3:} If $s$ sees $s'$ and $s' = t$ is the end, then we can pick any non level edge and flip it to $ss'$ to decrease the potential.
\end{itemize}
So now, suppose that $s$ sees $s'$, and the incoming edge of $s'\neq t$ is a level edge. Cases~4-6 follow different strategies depending on the type of $\out(s')$.

\begin{itemize}
    \item \textbf{Case~4:} If $\out(s')$ is a small diagonal, the idea is that we can rotate that diagonal, with $x = \succ(s')$ as an anchor, via the flips $s'x\to sx$ followed by $\inc(s)\to s's$. 
    We repeat applying this case until a level edge is created as illustrated in Figure~\ref{fig:inner-3-2}. The validity of the sequence comes from two loop invariants:
    \begin{itemize}
        \item $s,s'\in L_1$ are neighbors and see each other;
        \item $s'x$ is a small diagonal with $x=\succ(s')$,
    \end{itemize}
    and the termination is implied by the fact that after an iteration of the loop, the couple $(s, s')$ is shifted clockwise, so if the loop does not stop earlier, the diagonal $s'x$ will rotate until it becomes a level edge, when $s'$ becomes $\ccw(x)$. This loop can rotate that diagonal at most $O(n_1)$ times, meaning that the reduction step in this case takes $O(n_1)$ flips.
    
\begin{figure}[ht]
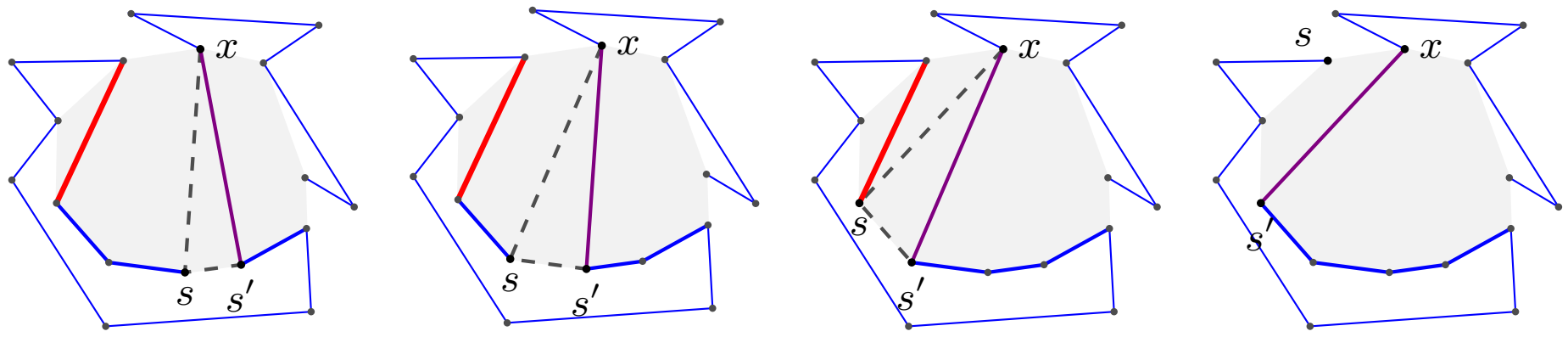

    \centering
    \begin{subfigure}[t]{0.23\textwidth}
        \centering
        \FigureEndpointsInCaseIIIdIIi
        \label{fig:inner-3-2_1}
    \end{subfigure}
    \hfill
    \begin{subfigure}[t]{0.23\textwidth}
        \centering
        \FigureEndpointsInCaseIIIdIIii
        \label{fig:inner-3-2_2}
    \end{subfigure}
    \hfill
    \begin{subfigure}[t]{0.23\textwidth}
        \centering
        \FigureEndpointsInCaseIIIdIIiii
        \label{fig:inner-3-2_3}
    \end{subfigure}
    \hfill
    \begin{subfigure}[t]{0.23\textwidth}
        \centering
        \FigureEndpointsInCaseIIIdIIiv
        \label{fig:inner-3-2_4}
    \end{subfigure}

    \caption{Rotation of a small diagonal in $O(n_1)$ flips until the potential decreases.}
    \label{fig:inner-3-2}
\end{figure}

    \item \textbf{Case 5:} If $\out(s')$ is a medium diagonal, the idea is to use first rotate the diagonal until the potential diminishes as illustrated in Figure~\ref{fig:inner-3-3-2}, or $x = \succ(s')$ does not see $s$ (Case~5.(a)), and then if the potential has not already be diminished, use the procedure \textsc{DiagObstrReduction} on the obstruction between $s$ and $x$ (Case~5.(b)).
    
    \begin{figure}[ht]
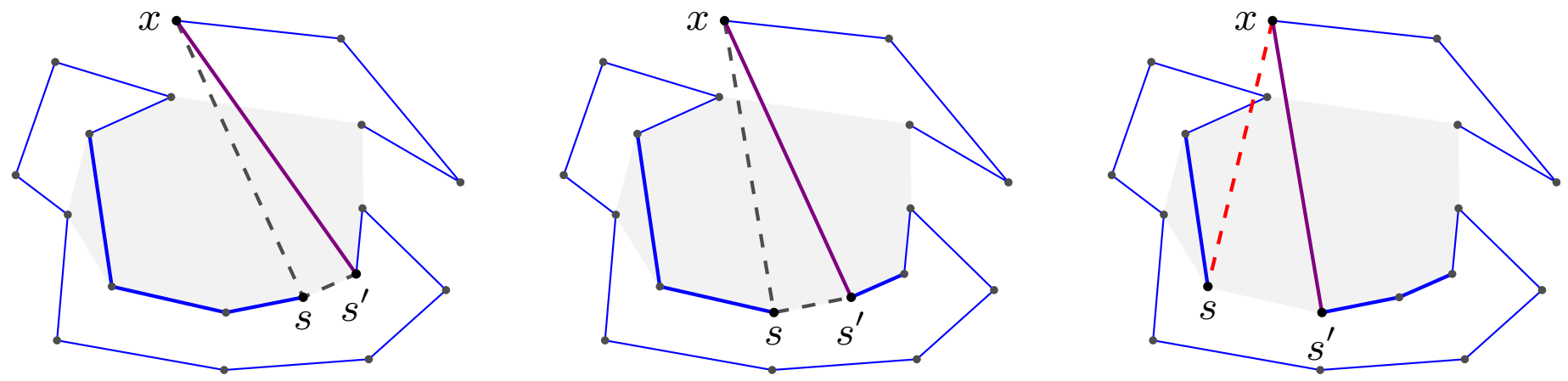

    \centering
    \begin{subfigure}[t]{0.3\textwidth}
        \centering
        \FigureEndpointsInCaseIIIdIIIdIIi
        \label{fig:inner-3-3-2_1}
    \end{subfigure}
    \hfill
    \begin{subfigure}[t]{0.3\textwidth}
        \centering
        \FigureEndpointsInCaseIIIdIIIdIIii
        \label{fig:inner-3-3-2_2}
    \end{subfigure}
    \hfill
    \begin{subfigure}[t]{0.3\textwidth}
        \centering
        \FigureEndpointsInCaseIIIdIIIdIIiii
        \label{fig:inner-3-3-2_3}
    \end{subfigure}

    \caption{Rotation of a medium diagonal in Case 5.(a) of \textsc{InnerReduction} in $O(n_1)$ flips.}
    \label{fig:inner-3-3-2}
\end{figure}
    
    The proof for the validity of Case~5.(a) is similar to Case~4, so let's focus on Case~5.(b). If $s$ does not see $x$, and $t$ is on the other side of the medium diagonal, we can show that the path verifies the premises of Lemma~\ref{lem:correct_diagobstr}. Let $u=s'$ and $v=x$.
    Then since $v$ does not see $s$, the path must enter the triangle $\triangle suv$ without crossing $uv$ (planarity) or $su$ (assumption). Thus the existance of an inner point $p\in\triangle suv$ the closest to $uv$. Then we can prove that $p$ sees $s$ and $v$.
    
    Also, we can observe that the prefix $P[s:u]$ must contain an outer point. Let $u'\in L_0$ be the last one of that prefix. Then the idea is that $P[v:u']$ separates the set of point in 2 parts, and the prefix must visit all the points of it's left side. With this observation and the fact that $s$ sees $u$, we can deduce that the path cannot visit any outer point before reaching $p$ without crossings. And similar planarity restrictions enforce the prefix $P[s:p]$ to visit a subset of points of $L_1$ in clockwise order, and the subpath $P[p:u']$ to contain at least all the points of $L_0$ between $v' =\ccw(v)$ and $u'$ in counterclockwise order. Therefore the premises of \textsc{DiagObstrReduction} are met and it can reduce the potential in $O(1)$ flips. An example of a path in this case is given in Figure~\ref{fig:inner-3-3-1}.

    \begin{figure}[ht]
       \begin{minipage}{0.48\textwidth}
         \centering
         \FigureEndpointsInCaseIIIdIIIdI
         \caption{Prefix structure of the path in Case 5.(b) of \textsc{InnerReduction}}
         \label{fig:inner-3-3-1}
       \end{minipage}\hfill
       \begin{minipage}{0.48\textwidth}
         \centering
         \FigureEndpointsInCaseIIIdIVdI
         \caption{Illustration of the dashed polygonal line separating the points in two parts in Case 6.(a) of \textsc{InnerReduction}}\label{fig:inner-3-4-1}
       \end{minipage}
    \end{figure}

    \item \textbf{Case 6:} If $\out(s')$ is a bridge, the natural goal is to remove it. We subdivise that case since the path can have different properties depending on the shape of the prefix.
    \begin{itemize}
        \item \textbf{Case 6.(a):} If $P[s:s']\not\subset L_1$
    To enforce stong properties on the path, suppose $s'$, $u=\succ(s')$ and $\succ(u)$ are in clockwise order (otherwise we can come back to this case with a single flip). Then let $x$ be the first outer point visited by the path. Then we can show that the polygonal line formed by $us's$ and $P[s:x]$ splits the points in half, with the suffix $P[u:t]$ on the left side, and the subpath $P[x:s']$ on the left as illustrated in Figure~\ref{fig:inner-3-4-1}. We can deduce from this structure that the prefix $P[s:u]$ must visit every outer point from $x$ to $u$ in clockwise order.
    So the last outer point visited before $u$ was $v = \ccw(u)$.
    Let $v' = \succ(v)$.

    If $s$ sees $v'$ (Case 6.(a).i), then the flip $vv'\to sv'$ transforms a bridge or medium diagonal into a small diagonal, and $s'u\to vu$ creates a level edge on the $L_0$ as illustrated in Figure~\ref{fig:inner-3-4-1-1}

    \begin{figure}[ht]
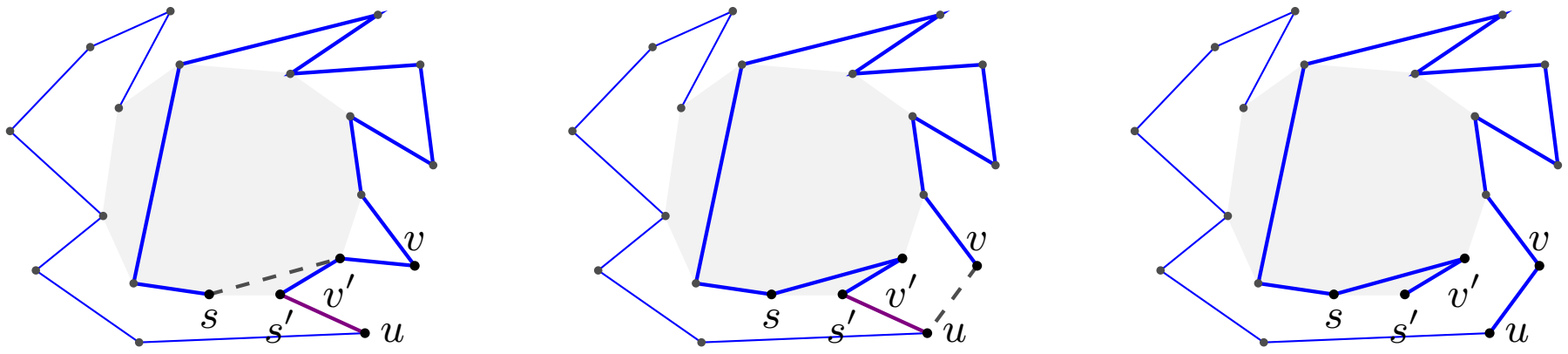

    \centering
    \begin{subfigure}[t]{0.3\textwidth}
        \centering
        \FigureEndpointsInCaseIIIdIVdIdIi
        \label{fig:inner-3-4-1-1_1}
    \end{subfigure}
    \hfill
    \begin{subfigure}[t]{0.3\textwidth}
        \centering
        \FigureEndpointsInCaseIIIdIVdIdIii
        \label{fig:inner-3-4-1-1_2}
    \end{subfigure}
    \hfill
    \begin{subfigure}[t]{0.3\textwidth}
        \centering
        \FigureEndpointsInCaseIIIdIVdIdIiii
        \label{fig:inner-3-4-1-1_3}
    \end{subfigure}

    \caption{Reduction of the potential in Case~6.(a).ii of \textsc{InnerReduction}}
    \label{fig:inner-3-4-1-1}
\end{figure}
    
    Otherwise (Case 6.(a).ii), there must be a small diagonal in $P[v':s']$ blocking the vision between $s$ and $v$. Let $pq$ be the last small diagonal of that subpath. Then $P[q:s']$ only contains level edges and $s$ sees $s'$ so the flip $pq\to sq$ is valid and does not change the potential.

    Now, let's show that applying the same procedure on the new path does not make a loop. Let $p' = \ccw(p)$. That point does not share an edge with $p$ because otherwise $P[v':p']$ would have had to cross $pq$. Therefore when applying this procedure, the choice of $p'$ for the neighbor of the start $p$ of the new path $P'$ is valid, and we can show that in $P'$, if the incoming edge of $p'$ is a level edge, its outgoing edge is a bridge and it sees $p$, then that bridge must be crossing the subpath $P'[p:p']$. Therefore Case 6 cannot be applied again in $P'$, so one of the other ones will. So in the worst case, the potential will be diminished in $O(n_1)$ flips. An illustration of that case is presented in Figure~\ref{fig:inner-3-4-1-2}.

    \begin{figure}[ht]
       \begin{minipage}{0.48\textwidth}
         \centering
         \begin{subfigure}[t]{0.48\textwidth}
            \centering
            \FigureEndpointsInCaseIIIdIVdIdIIi
            \label{fig:inner-3-4-1-2_1}
        \end{subfigure}
        \hfill
        \begin{subfigure}[t]{0.48\textwidth}
            \centering
            \FigureEndpointsInCaseIIIdIVdIdIIii
            \label{fig:inner-3-4-1-2_2}
        \end{subfigure}

    \caption{Reduction from Case 6.(a).ii to other cases of \textsc{InnerReduction}.}
    \label{fig:inner-3-4-1-2}
       \end{minipage}\hfill
       \begin{minipage}{0.48\textwidth}
         \centering
         \begin{subfigure}[t]{0.48\textwidth}
        \centering
        \FigureEndpointsInCaseIIIdIVdIIi
        \label{fig:inner-3-4-2_1}
    \end{subfigure}
    \hfill
    \begin{subfigure}[t]{0.48\textwidth}
        \centering
        \FigureEndpointsInCaseIIIdIVdIIii
        \label{fig:inner-3-4-2_2}
    \end{subfigure}

    \caption{Reduction from Case 6.(b) to other cases of \textsc{InnerReduction}.}
    \label{fig:inner-3-4-2}
       \end{minipage}
    \end{figure}

    \item \textbf{Case 6.(b)} Otherwise $P[s:s']\subset L_1$. In this last case, there is no immediate reduction possible in the general case, so we perform a flip to reduce that case to the other ones.
    Let $uv$ be the first small diagonal of $P[s:s']$. It must exist as $t\in L_1$ is not in $P[s:s']$. Then $u$ has neighbor $u'$ not seen by $s$, and $u'$ cannot be connected to $u$ because otherwise $P[u':s']$ would need to cross $uv$. 
    
    So after the flip $uv\to sv$, $u$ becomes the start of the new path $P'$ and $u'$ is a valid choice for the neighbor of $u$ in the recursive call of the procedure for $P'$. Also, we can be sure that $P'$ will not fall again into Case~6.(b) since $u\notin P[s:s']$, so $u\notin P'[u:s']$ and therefore $P'[u:u']\not\subset L_1$. So in the worst case, the potential will be diminished in $O(n_1)$ flips. This reduction of case is represented in Figure~\ref{fig:inner-3-4-2}.
    \end{itemize}
\end{itemize}

Therefore each case uses at most $O(n_1)$ flips to reduce the potential.
\end{proof}

\subsection{Obstruction procedures and proof of Lemmas~\ref{lem:correct_obstr} and~\ref{lem:correct_diagobstr}}\label{sec:obstr-reduction}
\textbf{Procedure} \textsc{ObstrReduction}.

\textit{Input:}
a path $P$ whose start $s\in L_1$ does not see one of its neighbor $q\in L_1$. 

Assume $q=\ccw(s)$ by symmetry.

Let $uv$ be the closest edge to $s$ intersecting $sq$,
with $u=\pred(v)$.

\begin{enumerate}
\item If $v$ sees $s$, perform $uv \longrightarrow sv.$

\item Else follow the procedure \textsc{DiagObstrReduction}.
\end{enumerate}

\noindent
\textbf{Procedure} \textsc{DiagObstrReduction}.
    \textit{Input:} A path $P$ starting at $s\in L_1$ such that there is a medium or big diagonal $uv$, with $v\in L_0$, intersecting the convex hull of the inner layer next to $s$ and another vertex $p$. Suppose also that $p$ is inside the triangle $\triangle suv$, and that the prefix $P[s:p]$ visits a subset of $L_1$ in circular order from $s$ to $p$, and the subpath $P[p:u']$ visits all the points between $s$ and $p$ skipped by the prefix (without particular order), and all the outer points in order from $v'$ (the first neighbor of $v$ visited by after $p$) to $u'$ (the last outer point visited before $u$).
    
      Suppose without loss of generality that the crossing of the diagonal $uv$ is counterclockwise to $s$.
      Let $s'=\pred(p)$, $v'=\ccw(v)$ and $p'=\ccw(p)$.

      \begin{enumerate}

      \item If $\succ(p)\in L_1$, perform
            \[
            s'p\to sp,
            \qquad
            p\,\succ(p)\to s'\succ(p),
            \qquad
            uv\to pv.
            \]

      \item Else if $p'\in P[s:p]$, perform
              \[
              s'p\to sp,
              \qquad
              pv'\to s'v',
              \qquad
              uv\to pv.
              \]

        \item Else if $p'$ does not see $p$:

              Let $ab$, with $b=\succ(a)$,
              be the closest edge intersecting $pp'$
              and perform
              \[
              s'p\to sp,
              \qquad
              ab\to s'b.
              \]

        \item Otherwise, perform
              \[
              s'p\to sp.
              \]

              \begin{enumerate}

              \item If $\succ(p')\in L_1$, let
                    $y=\succ(p')$ and perform
                    \[
                    p'y\to s'y,
                    \qquad
                    v'p\to p'p,
                    \qquad
                    uv\to vv'.
                    \]

              \item Else if
                    $\pred(p'),\succ(p')\in L_0$,
                    
                    Let $x=\pred(p')$ and $y=\succ(p')$
                    and perform
                    \[
                    xp'\to s'p',
                    \qquad
                    p'y\to xy.
                    \]

              \item Otherwise, $\succ(p')\in L_0$ and $\pred(p')\in L_1$.
              \begin{enumerate}
                  \item if $s'$ does not see
                          $\cw(s')$, recursively apply \textsc{ObstrReduction} on the obstruction between $s'$ and $\cw(s')$.

            \item Else if the incoming edge of
                  $\cw(s')$ is not level, perform
                  \[
                  \operatorname{in}(\cw(s'))
                  \to s'\cw(s').
                  \]

            \item Otherwise, perform
                  \[
                  \operatorname{in}(\cw(s'))
                  \to s'\cw(s'),
                  \qquad
                  \operatorname{in}(p')
                  \to p'\cw(\cw(s')).
                  \]
              \end{enumerate}

              \end{enumerate}

      \end{enumerate}

\begin{proof}[Lemma~\ref{lem:correct_obstr}]
    Let $P$ be an $st$-path whose start $s\in L_1$ does not see one of its neighbor $q$, and assume $q = \ccw(s)$ by symmetry.
    Let $uv$ be the closest edge to $s$ intersecting $sq$. Then $uv$ is either a medium diagonal or a big diagonal, so it has a lot of weight. Therefore the goal is to remove it. We can observe that because of the definition of $uv$, $s$ sees either $u$ or $v$.

    \begin{itemize}
        \item \textbf{Case 1.} If $s$ sees $v$, then the flip $uv\to sv$ is legal and we can show by case exhaustion that no matter the type of $uv$ initially, that flip decreases the potential. The different subcases are illustrated in Figure~\ref{fig:obstr-1to3}.
        
        \begin{figure}[ht]
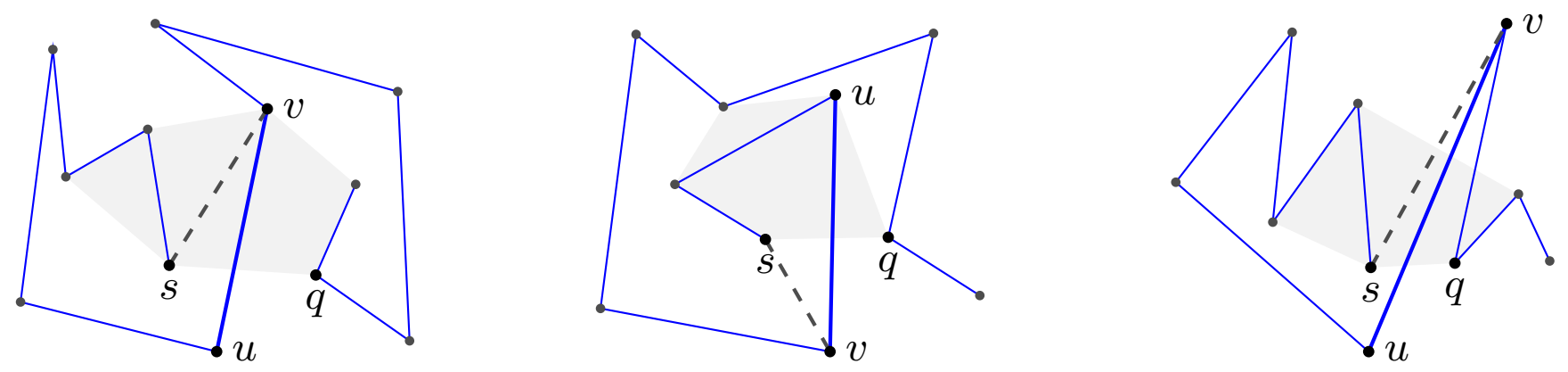

    \centering

    \begin{subfigure}[t]{0.3\textwidth}
        \centering
        \FigureObstrCaseI
        \caption{$v$ is an inner point}
        \label{fig:obstr-1}
    \end{subfigure}
    \hfill
    \begin{subfigure}[t]{0.3\textwidth}
        \centering
        \FigureObstrCaseII
        \caption{$v$ is an outer point next to $s$}
        \label{fig:obstr-2}
    \end{subfigure}
    \hfill
    \begin{subfigure}[t]{0.3\textwidth}
        \centering
        \FigureObstrCaseIII
        \caption{$v$ is an outer point far from $s$}
        \label{fig:obstr-3}
    \end{subfigure}

    \caption{Immediate potential reduction in case 1 of \textsc{ObstrReduction}}
    \label{fig:obstr-1to3}
\end{figure}

        \item \textbf{Case 2.} Otherwise, $s$ does not see $v$ but $s$ sees $u$, so we can deduce that the diagonal $uv$ must be a big diagonal, otherwise, $s$ would have been able to see $v$. 
        We can now show that the path has the strong structural properties that enable the use of the procedure \textsc{DiagObstrReduction}. 

        Since $s$ does not see $v$, the path must enter the triangle $\triangle suv$ without crossing $uv$ (planarity) or $su$ (assumption). so that triangle contains at least 1 point of $L_1$ inside of it. Let $p$ be the closest point to $uv$ inside that triangle. Then we can prove that there is no other point inside the concave quadrilateral $suvp$ or that would contradict the definition of $p$, therefore $p$ sees both $s$ and $v$.

        Now, since $uv$ is a big diagonal, it splits the point set in two parts, and it particular, it must separate the start $s$ from the end $t$ because otherwise an entire side could not be visited without crossings. This means that the prefix $P[s:u]$ must visit the entire left side of $uv$. Moreover, since the quadrilateral $suvp$ cannot be crossed by the path, we can deduce that the prefix $P[s:p]$ cannot visit any point of $L_0$ without introducing crossings or violating the definition of $uv$ or $p$.
        Therefore $P[p:u]$ must visit at least all the outer points from $\ccw(v)$ to $u$. With the same argument, we can observe that $P[s:p]$ must visit that subset of $L_1$ in clockwise order and all the outer points of the left side are visited in counterclockwise order in $P[p:u]$. 
        The structure of the prefix of that path is illustrated in Figure~\ref{fig:obstr-prefix-structure}
        
        Therefore the premises of the procedure \textsc{DiagObstrReduction} are met and it can reduce the potential in $O(1)$ flips. 
        \begin{figure}[ht]
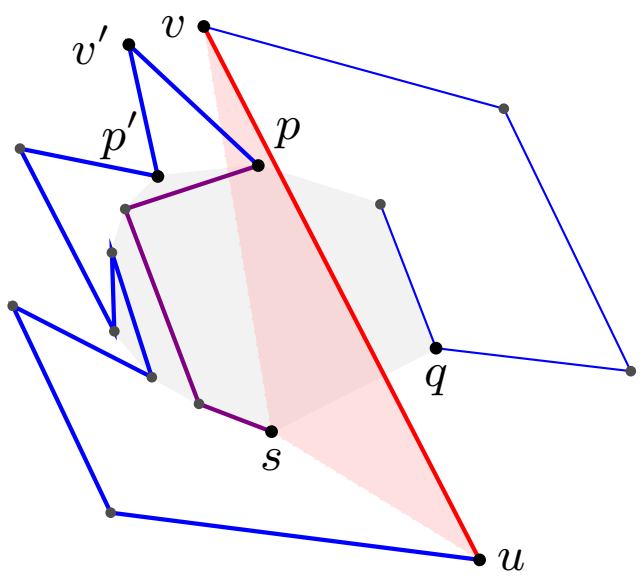

            \centering
            \FigurePrefixStructure
            \caption{Illustration of the particular prefix the path has in Case 2 of \textsc{ObstrReduction}}
            \label{fig:obstr-prefix-structure}
        \end{figure}
    \end{itemize}

    Therefore in both cases, the potential of the path can be reduced in $O(1)$ flips.\qed
\end{proof}

\begin{proof}[Lemma \ref{lem:correct_diagobstr}]
The main goal of this procedure is to remove the diagonal $uv$ by flipping it to $pv$, or to remove any other heavy weight edge encountered in the prefix. This procedure does it by successively eliminating simple cases to constrain the shape of the path until it forces the appearance of a medium diagonal in the prefix that is easy to remove.

Let $s' = \pred(p)$. A first easy case is when the successor of $p$ is an inner point.
\begin{itemize}
    \item \textbf{Case 1.} If $\succ(p)\in L_1$, we can make $p$ the new starting point with 2 flips: $s'p\to sp$ followed by $p\,\succ(p)\to s'\succ(p)$. These flips are legal since $p$ sees $s$ and $s'p\succ(p)$ is an empty triangle of $L_1$ where two of the borders cannot be crossed by planarity, so $s'$ sees $\succ(p)$. After these two flips, the potential may have increased by 1 at most, However, the last flip $uv\to pv$ transform a medium or big diagonal into a bridge, diminishing the potential by at least 2. So at the end of that sequence, the potential has decreased by at least 1.

    This flip sequence is illustrated in Figure~\ref{fig:obstr-4-1}
\begin{figure}[ht]
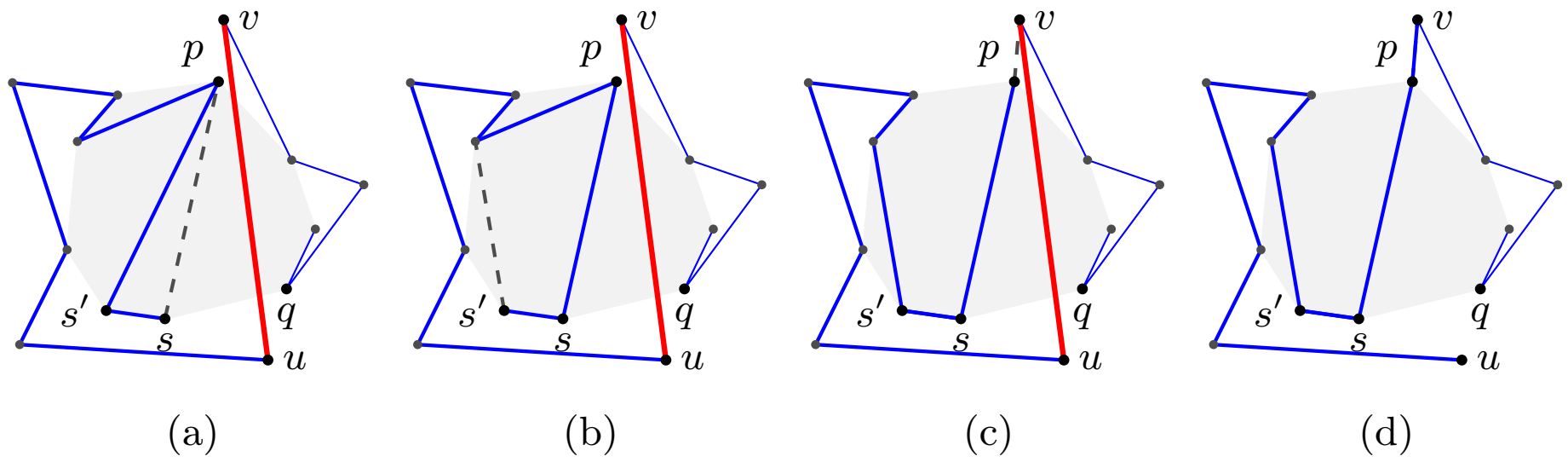

    \centering
    \begin{subfigure}[t]{0.24\textwidth}
        \centering
        \FigureObstrCaseIVdIi
        \caption{}
        \label{fig:obstr-4-1_1}
    \end{subfigure}
    \hfill
    \begin{subfigure}[t]{0.24\textwidth}
        \centering
        \FigureObstrCaseIVdIii
        \caption{}
        \label{fig:obstr-4-1_2}
    \end{subfigure}
    \hfill
    \begin{subfigure}[t]{0.24\textwidth}
        \centering
        \FigureObstrCaseIVdIiii
        \caption{}
        \label{fig:obstr-4-1_3}
    \end{subfigure}
    \hfill
    \begin{subfigure}[t]{0.24\textwidth}
        \centering
        \FigureObstrCaseIVdIiv
        \caption{}
        \label{fig:obstr-4-1_4}
    \end{subfigure}

    \caption{Reduction of potential in Case~1 of \textsc{DiagObstrReduce}}
    \label{fig:obstr-4-1}
\end{figure}    
\end{itemize}

Suppose now that $\succ(p)\in L_0$. Since by assumption the subpath $P[p:u]$ visits all the outer points from $v' = \ccw(v)$ to $u$ in counterclockwise order, we can deduce that $\succ(p) = v'$. Let $p' = \ccw(p)$. The case analysis is now refined based on the position of $p'$ in the path
\begin{itemize}
    \item \textbf{Case 2.} If $p'\in P[s:p]$, then $p' = s'$ since by assumption on $P[s:p]$, that prefix visit points of $L_1$ in clockwise order. The sequence of flips we perform in this case is the same as in Case 1, except this time the second flip is legal because $v'$ and $p'$ see each other. Indeed, if that was not the case, the convexity of $L_1$ would prevent the quadrilateral $suvp$ from being concave, which would mean that $s$ would see $v$, contradicting the assumption made on the initial path. 

    So at the end of that sequence (illustrated in Figure~\ref{fig:obstr-4-2-1}, the potential has decreased by at least 1 in $O(1)$ flips.
    \begin{figure}[ht]
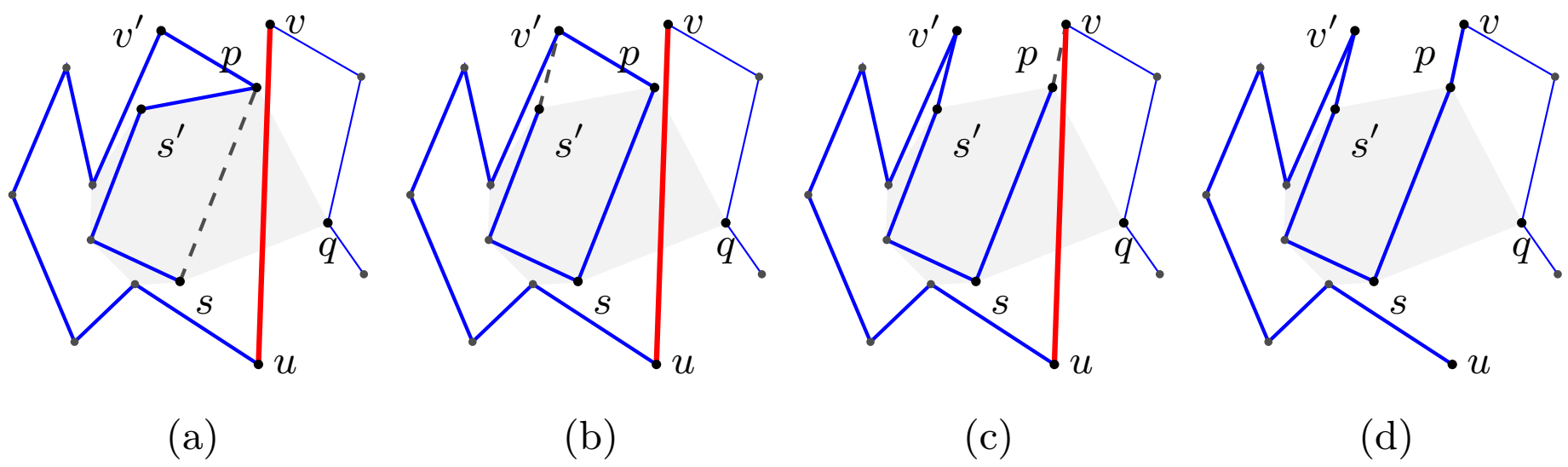

    \centering

    \begin{subfigure}[t]{0.24\textwidth}
        \centering
        \FigureObstrCaseIVdIIdIi
        \caption{}
        \label{fig:obstr-4-2-1_1}
    \end{subfigure}
    \hfill
    \begin{subfigure}[t]{0.24\textwidth}
        \centering
        \FigureObstrCaseIVdIIdIii
        \caption{}
        \label{fig:obstr-4-2-1_2}
    \end{subfigure}
    \hfill
    \begin{subfigure}[t]{0.24\textwidth}
        \centering
        \FigureObstrCaseIVdIIdIiii
        \caption{}
        \label{fig:obstr-4-2-1_3}
    \end{subfigure}
    \hfill
    \begin{subfigure}[t]{0.24\textwidth}
        \centering
        \FigureObstrCaseIVdIIdIiv
        \caption{}
        \label{fig:obstr-4-2-1_4}
    \end{subfigure}

    \caption{Reduction of potential in Case~2 of \textsc{DiagObstrReduction}}
    \label{fig:obstr-4-2-1}
\end{figure}
    
    \item \textbf{Case 3.} If $p'\notin P[s:p]$ and $p'$ does not see $p$,
    then there is a medium diagonal blocking the view between those points. Let $ab$ be the closest edge to $p$ intersecting $pp'$. Then we can show that necessarily $a\in L_0$ and $b\in L_1$, otherwise there would need to be some crossings in the path. We can also show that $s'$ sees $b$ since the triangle $\triangle s'pb$ is empty and the path cannot cross $s'p$ (planarity) nor can it cross $bp$ (definition of $ab$). So after the first neutral flip $s'p\to sp$, the flip $ab\to s'b$ reduced the potential by at least 2 as illustrated in Figure~\ref{fig:obstr-4-2-2}.
    \begin{figure}[ht]
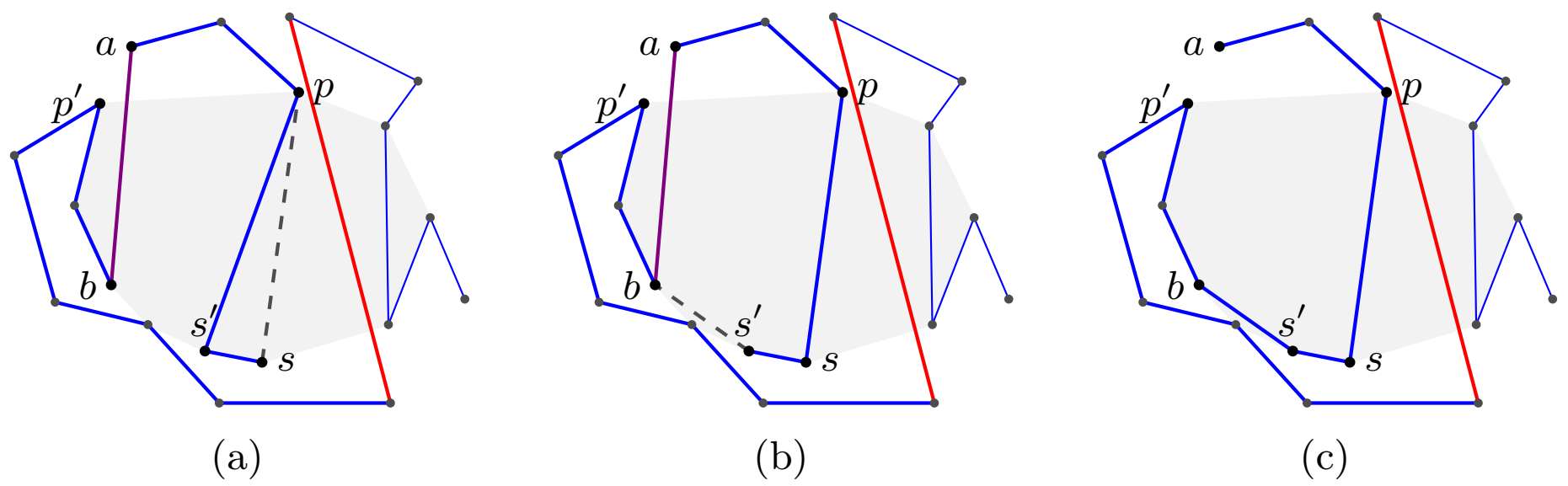

    \centering
    \begin{subfigure}[t]{0.3\textwidth}
        \centering
        \FigureObstrCaseIVdIIdIIi
        \caption{}
        \label{fig:obstr-4-2-2_1}
    \end{subfigure}
    \hfill
    \begin{subfigure}[t]{0.3\textwidth}
        \centering
        \FigureObstrCaseIVdIIdIIii
        \caption{}
        \label{fig:obstr-4-2-2_2}
    \end{subfigure}
    \hfill
    \begin{subfigure}[t]{0.3\textwidth}
        \centering
        \FigureObstrCaseIVdIIdIIiii
        \caption{}
        \label{fig:obstr-4-2-2_3}
    \end{subfigure}

    \caption{Reduction of potential in Case~3 of \textsc{DiagObstrReduction}.}
    \label{fig:obstr-4-2-2}
\end{figure}
\end{itemize}

Now suppose $p'$ sees $p$ and $p'\notin P[s:p]$ (Case 4.), i.e. $p'\in P[v':u]$.
Then like all the previous cases, we start flipping $s'p\to sp$ to form a new path $P'$ as illustrated in Figure~\ref{fig:obstr-4-2-3}. In this case, we know that $s'$ is not a cyclic neighbor of $p$ in the path, so $s'p$ is a small diagonal, just like $sp$. Therefore $\Phi(P) = \Phi(P')$.

\begin{figure}[ht]
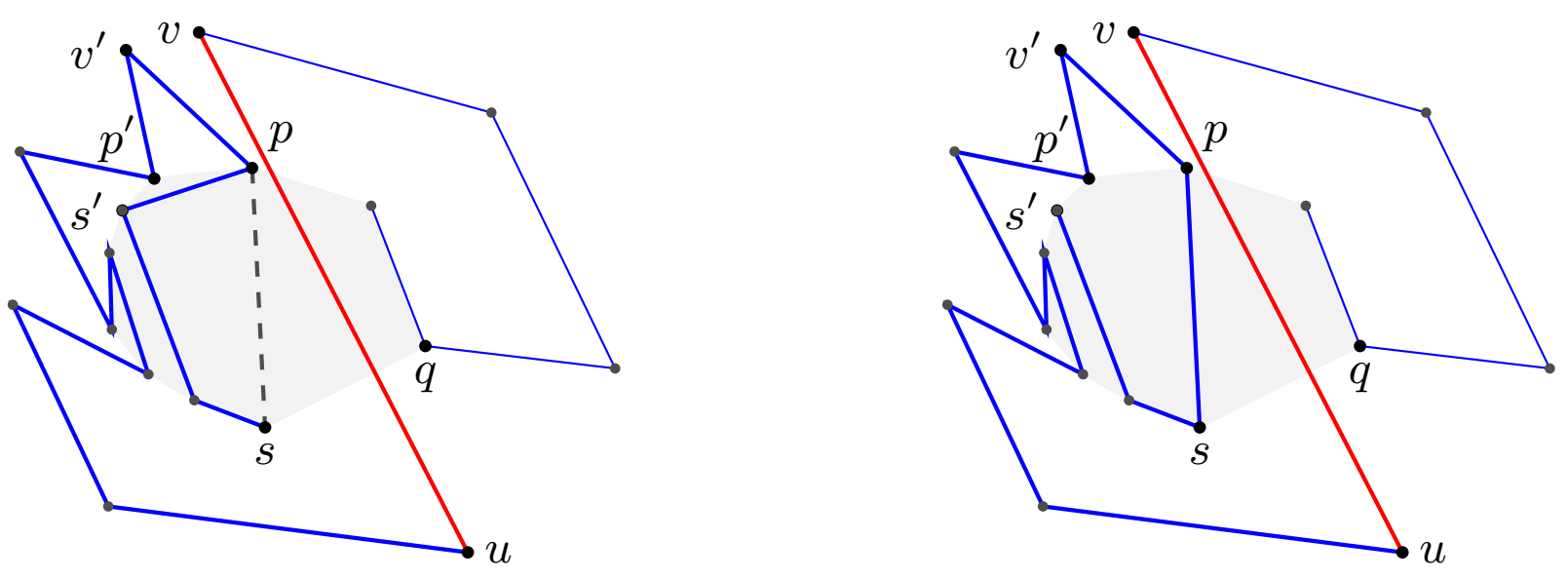

    \centering
    \begin{subfigure}[t]{0.48\textwidth}
        \centering
        \FigureObstrCaseIVdIIdIIIi
        \caption{The original path $P$.}
        \label{fig:obstr-4-2-3_1}
    \end{subfigure}
    \hfill
    \begin{subfigure}[t]{0.48\textwidth}
        \centering
        \FigureObstrCaseIVdIIdIIIii
        \caption{The new path $P'$.}
        \label{fig:obstr-4-2-3_2}
    \end{subfigure}

    \caption{First flip of Case~4 of \textsc{DiagObstrReduction}.}
    \label{fig:obstr-4-2-3}
\end{figure}

We refine again the case analysis based on the position of the predecessor and the successor of $p'$. We can observe that 2 cases can be solved easily.

\begin{itemize}
    \item \textbf{Case 4.(a)}: If $\succ(p')\in L_1$, This case resembles the flip sequence from Case 1: with the same empty triangles arguments we can deduce the required visibility to prove that all the flips performed in this case are legal. The potential could stay the same after the first 2 flips but decreases by at least 3 with the last one as illustrated in Figure~\ref{fig:obstr-4-2-3-1}.
    \begin{figure}[ht]
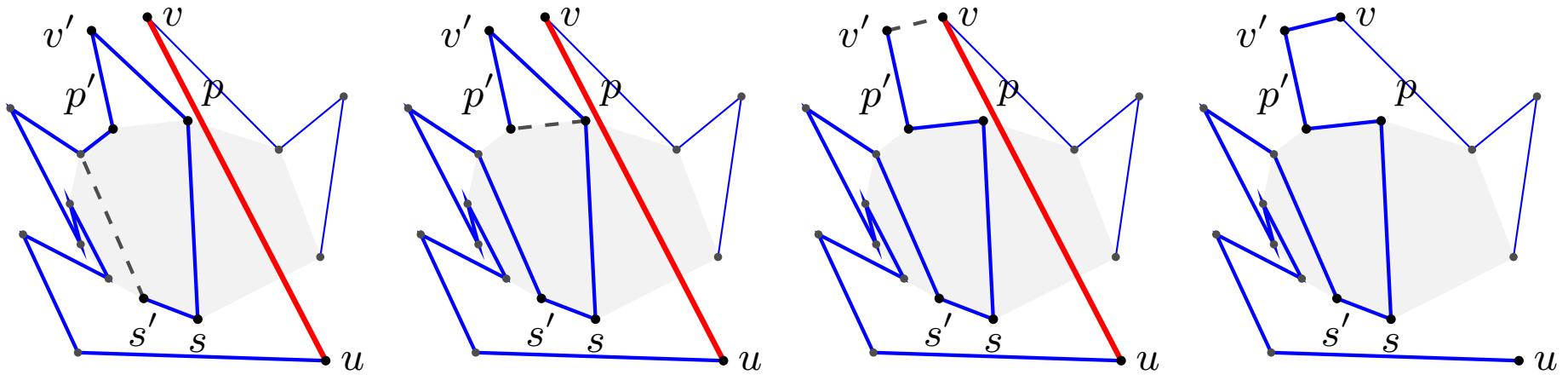

    \centering
    \begin{subfigure}[t]{0.24\textwidth}
        \centering
        \FigureObstrCaseIVdIIdIIIdIi
        \caption{}
        \label{fig:obstr-4-2-3-1_1}
    \end{subfigure}
    \hfill
    \begin{subfigure}[t]{0.24\textwidth}
        \centering
        \FigureObstrCaseIVdIIdIIIdIii
        \caption{}
        \label{fig:obstr-4-2-3-1_2}
    \end{subfigure}
    \hfill
    \begin{subfigure}[t]{0.24\textwidth}
        \centering
        \FigureObstrCaseIVdIIdIIIdIiii
        \caption{}
        \label{fig:obstr-4-2-3-1_3}
    \end{subfigure}
    \hfill
    \begin{subfigure}[t]{0.24\textwidth}
        \centering
        \FigureObstrCaseIVdIIdIIIdIiv
        \caption{}
        \label{fig:obstr-4-2-3-1_4}
    \end{subfigure}

    \caption{Reduction of potential in Case~4.(a) of \textsc{DiagObstrReduction}.}
    \label{fig:obstr-4-2-3-1}
\end{figure}

    \item \textbf{Case 4.(b)} If $\succ(p')\in L_0$ and $\pred(p')\in L_0$, then by the assumption on the path, we know that $x = \pred(p')$ is the clockwise neighbor of $y = \succ(p')$. In this case we can just connect them in 2 flips. By the same empty triangle argument we know that $s'$ sees $p'$, and since they are outer layer neighbor, we know that $x$ sees $y$, thus the validity of both flips. Moreover, that sequence of flips transforms two bridges into a level edge and a small diagonal, so the potential decreases by 1 as illustrated in Figure~\ref{fig:obstr-4-2-3-2}.

\begin{figure}[ht]
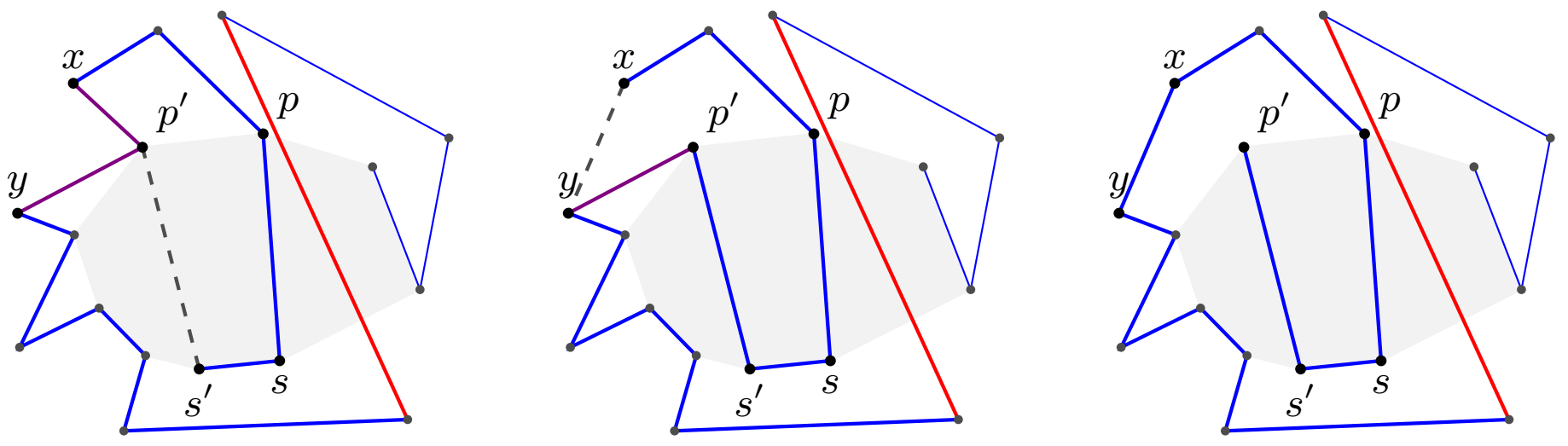

    \centering
    \begin{subfigure}[t]{0.3\textwidth}
        \centering
        \FigureObstrCaseIVdIIdIIIdIIi
        \caption{}
        \label{fig:obstr-4-2-3-2_1}
    \end{subfigure}
    \hfill
    \begin{subfigure}[t]{0.3\textwidth}
        \centering
        \FigureObstrCaseIVdIIdIIIdIIii
        \caption{}
        \label{fig:obstr-4-2-3-2_2}
    \end{subfigure}
    \hfill
    \begin{subfigure}[t]{0.3\textwidth}
        \centering
        \FigureObstrCaseIVdIIdIIIdIIiii
        \caption{}
        \label{fig:obstr-4-2-3-2_3}
    \end{subfigure}

    \caption{Reduction of potential in Case~4.(b) of \textsc{DiagObstrReduction}.}
    \label{fig:obstr-4-2-3-2}
\end{figure}
\end{itemize}

Now suppose $\succ(p')\in L_0$ and $\pred(p')\in L_1$. In this case, the outgoing edge of $p'$ is so constrained that it is forced to be a medium diagonal as we will show in the relevant subcase. Therefore, we shift the objective from flipping $uv$ to flipping $\out(p')$. However, the edge is in the wrong direction to be flipped directly. Therefore we use $\cw(s')$ to change the orientation of that edge to flip it (Case 4.(c).iii), if the potential cannot be reduced immediately (Case 4.(c).i-ii).

\begin{itemize}
    \item \textbf{Case 4.(c).i} If $s'$ does not see $\cw(s')$, because of the structure of the prefix assumed in the hypotheses of the lemma, that blocking edge cannot be a big diagonal, therefore it is a medium one, so applying procedure \textsc{ObstructionReduce} will reduce the potential with its case~1 in 1 flip.

    \item \textbf{Case 4.(c).ii} If $s'$ sees $\cw(s')$ and $\inc(\cw(s'))$ is not a level edge, then the flip $\inc(\cw(s'))\to s'\cw(s)$ creates a level edge, reducing the potential.

    \item \textbf{Case 4.(c).iii} if $s'$ sees $\cw(s')$ and $\inc(\cw(s'))$ is a level edge, then the flip $\inc(\cw(s'))\to s'\cw(s)$ makes $\cw(\cw(s'))$ the new start, and reverses the order of the path in $P'[s':\cw(s')]$ as displayed in Figure~\ref{fig:obstr-4-2-3-3-3}.

    The second flip is valid since after the first flip, $p'$ sees $s'$ which is connected to $\cw(s')$ which sees $\cw(\cw(s'))$, and all these points form an empty quadrilateral, so $p$ sees $\cw(\cw(s'))$.

    \begin{figure}[ht]
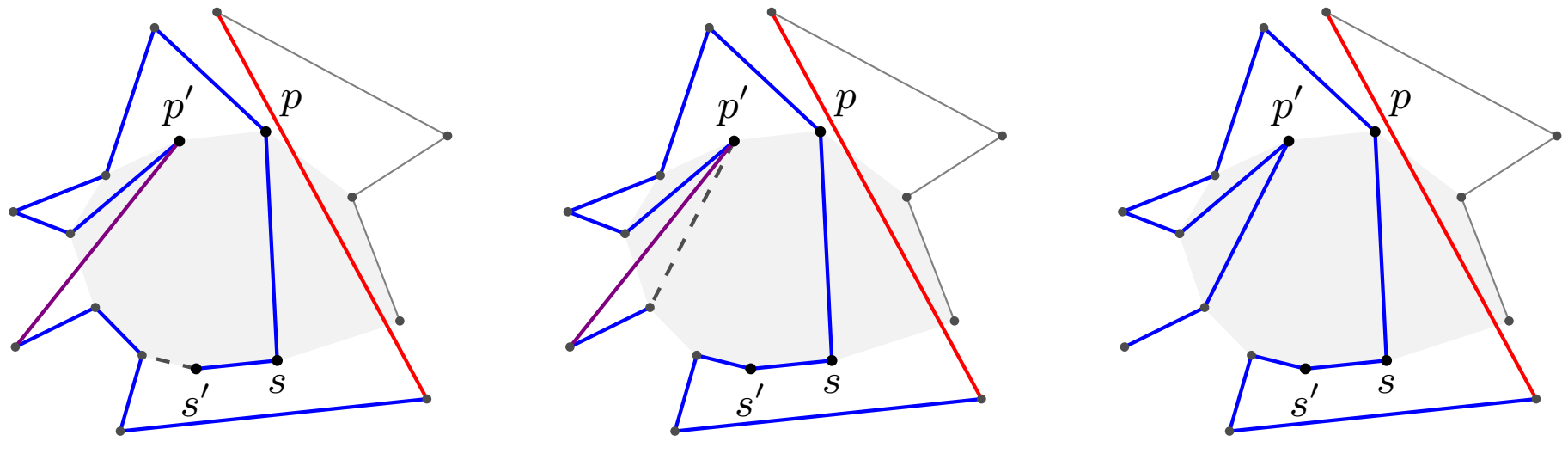

    \centering
    \begin{subfigure}[t]{0.3\textwidth}
        \centering
        \FigureObstrCaseIVdIIdIIIdIIIdIIIi
        \caption{}
        \label{fig:obstr-4-2-3-3-3_1}
    \end{subfigure}
    \hfill
    \begin{subfigure}[t]{0.3\textwidth}
        \centering
        \FigureObstrCaseIVdIIdIIIdIIIdIIIii
        \caption{}
        \label{fig:obstr-4-2-3-3-3_2}
    \end{subfigure}
    \hfill
    \begin{subfigure}[t]{0.3\textwidth}
        \centering
        \FigureObstrCaseIVdIIdIIIdIIIdIIIiii
        \caption{}
        \label{fig:obstr-4-2-3-3-3_3}
    \end{subfigure}

    \caption{Reduction of potential in Case~4.(c).iii of \textsc{DiagObstrReduction}.}
    \label{fig:obstr-4-2-3-3-3}
\end{figure}

    Now to prove that the potential reduces, we must prove that in $P'$, the outgoing edge of $p'$ is a medium diagonal, and that after the first flip, the order of the path in $p'$ is reversed, meaning that that medium diagonal is flipped for a small one in the second flip.

    Since $p' \in L_1$ and $\succ(p')\in L_0$, the outgoing edge of $p'$ is either a bridge or a medium diagonal. Suppose for contradiction that it is a bridge, and consider the simple loop $\Gamma = vp \cup pp'\cup p'\succ(p')\cup \gamma_{v,\succ(p')}$ where $\gamma_{v,\succ(p')}$ lies outside $\conv(S)$ except at its endpoints $v$ and $\succ(p')$. By the Jordan curve theorem, $\Gamma$ separates the plane into two regions. Since $p'\succ(p')$ is a bridge and $p'$ is a neighbor of $p$, the region containing $v'=\ccw(v)$ contains no point of $L_1$ other than $p$ and $p'$ at its border, so in particular, it does not contain $\pred(p')$. Therefore $P'[v':\pred(p')]$ must cross $\Gamma$, but it cannot cross $vp$ and $pp'$ (assumption), it cannot cross $p'\succ(p')$ (planarity) and it cannot cross $\gamma_{v, \succ(p')}$ (outside the hull). This contradiction shows that $\out(p')$ is a medium diagonal.

    We can also show with a different loop that $p'$ must be visited before $\cw(s')$ in $P'$, therefore after the first flip, the medium diagonal of $p'$ becomes its incoming edge and the potential diminishes by at least 2 in the second flip.
\end{itemize}

So in any case, the potential diminishes by at least 1 in $O(1)$ flips.\qed
\end{proof}

\section{Proof of Lemma~\ref{lem:not-too-many-hard-cases}}\label{sec:not-too-many-hard-cases}
Using Lemma~\ref{P2} alone, we would get a flip complexity of $O(n_1(n_0+n_1))$ in the worst case. In this section we prove that the expensive cases where we may need $\Omega(n_1)$ flips in a reduction step does not happen too often. In fact, we show that at most $O(n_1)$ reduction steps may need strictly more than 1 flip to be performed.

First, we can observe that the number of bridges $B(P)$, the number of small diagonals $S(P)$, and the number of medium diagonals $M(P)$ are all bounded by $2n_1$ as these edge types all have an extremity in $L_1$. Therefore, we just need to show that most of the big diagonals can be removed in 1 flip throughout the reduction process.

Let $\mathcal I$ be the set of boundary segments of $\conv(L_1)$ induced
by the vertices of $L_1$.

Let $P$ be a path. For $A,B\in\mathcal I$, let $G_P(A,B)$ denote the set of big diagonals whose two intersections with $\conv(L_1)$ belong respectively to the intervals $A$ and $B$.
We also denote by $G_P(\emptyset)$ the set of big diagonals that do not intersect $\conv(L_1)$, so that $G_P(\emptyset)\cup\bigcup_{A,B\in\mathcal I}G_P(A,B)$ forms a partition of the set of big diagonals of $P$. Because of the planarity of the path, the support of that partition is $O(n_1)$ (Lemma~\ref{lem:number-groups}).

\begin{lemma}\label{lem:number-groups}
For any given path $P$, the number of non-empty sets $G_P(A,B)$ is $O(n_1)$.
\end{lemma}

\begin{proof}
Associate to each non-empty set a representative big diagonal.
Since big diagonals of a plane path are pairwise non-crossing, these
representatives form a family of non-crossing chords of dual of the convex polygon $\conv(L_1)$.

The graph whose vertices are the segments of $\conv(L_1)$ (i.e. $\mathcal I$) and whose edges correspond to non-empty sets is therefore outerplanar.
Since $|\mathcal I|=n_1$, it contains only $O(n_1)$ edges.
Hence only $O(n_1)$ sets are non-empty.
\end{proof}

This partition of the big diagonals enables the following observation.

\begin{observation}\label{obs:GAB}
    Let $P$ be a path. As soon as one big diagonal of $G \in \{G_P(\emptyset)\}\cup\{G_P(A,B), A,B\in\mathcal I\}$ as been flipped, we can transform every remaining diagonal of $G$ to level edges using one flip per diagonal.
\end{observation}
\begin{proof}
    Suppose $G = G_P(A, B)\neq \emptyset$, for $A,B\in \mathcal I$, and suppose we are about to flip one diagonal of $G$. We can order them by the order in which they appear in the path (that would correspond to left-to-right or right-to-left if they were vertical).
    Let $uv$ be the first and $xy$ be last diagonal of $G$. 
    
    Since each big diagonal separates the point set in two parts, we can deduce that they also separate the endpoints (planarity and Hamiltonicity). Applying this principle on $uv$ and $xy$, we can deduce that there is no endpoint in the region of the plane between the lines $uv$ and $xy$. So no other diagonal from $G$ can be visible by the endpoints, therefore, by symmetry we can suppose that $uv$ is the diagonal we are about to flip, and that $u$ will become the next starting point.

    By definition of $G_P(A,B)$, there is no inner point in the region between the lines $uv$ and $xy$, so the vertices of the subpath $P[u:y]$ are in convex position. This forces a particular structure where each successive diagonal are connected by level edges, and as soon as $uv$ has been flipped, so that $u$ becomes an endpoint, we can successively flip all the diagonals in order to level edges like in the convex case.
    
    The proof is the similar when $G=G_P(\emptyset)$.\qed
\end{proof}

Therefore, according to Observation~\ref{obs:GAB}, at most $1+|G_P(A,B)\neq \emptyset| = O(n_1)$ big diagonals may need more than 1 flip in the reconfiguration process. We can now prove Lemma~\ref{lem:not-too-many-hard-cases}.

\begin{proof}[Lemma \ref{lem:not-too-many-hard-cases}]
    According to Observation~\ref{obs:GAB}, at most $B(P) + S(P) + 3M(P) + 4(1+|G_P(A,B)\neq \emptyset|) = O(n_1)$ reduction steps may require more than 1 flip. \qed
\end{proof}

\begin{credits}
\subsubsection{Acknowledgments and Disclosure of Interests.} Use of AI: Generative AI tools were used for literature search, language polishing, manuscript formatting, and improving the exposition of the paper. They were also used as an auxiliary aid to review the presentation of proofs and identify possible inconsistencies. Generative AI tools were not used to generate the scientific results, proofs, algorithms, or conclusions of this work. All AI-assisted content was reviewed and verified by the authors, who take full responsibility for the final manuscript. The authors have no competing interests to declare that are relevant to the content of this article.
\end{credits}

\bibliographystyle{splncs04}
\bibliography{dgmm}
\end{document}